\documentclass[fleqn,usenatbib]{mnras}

\usepackage{newtxtext,newtxmath}

\usepackage[utf8]{inputenc}

\usepackage[T1]{fontenc}
\usepackage{ae,aecompl}
\usepackage{ulem}

\usepackage{float}
\usepackage{graphicx}
\usepackage{amsmath}
\usepackage{multirow}
\usepackage{verbatim}
\usepackage{url}

\defcitealias{Cialone2018}{C18}

\title[Dynamical state and morphology of galaxy clusters]{The Three Hundred Project: Dynamical state of galaxy clusters and morphology from multi-wavelength synthetic maps}

\author[F. De Luca et al.]{\parbox{\textwidth}{
Federico De Luca,$^{1,2}$\thanks{E-mail: federico.deluca@roma2.infn.it}
Marco De Petris,$^{2}$
Gustavo Yepes,$^{3}$
Weiguang Cui,$^{4}$
Alexander Knebe,$^{3,5}$
Elena Rasia $^{6,7}$
}
\vspace{0.4cm}
\\
\parbox{\textwidth}{
$^{1}$Dipartimento di Fisica, Università di Roma “Tor Vergata”, Via della Ricerca Scientifica 1, 00133 Roma, Italy \\
$^{2}$Dipartimento di Fisica, Sapienza Università di Roma, Piazzale Aldo Moro 5, 00185 Roma, Italy \\
$^{3}$Departamento de Física Teórica and CIAFF, Módulo 8, Facultad de Ciencias, Universidad Autónoma de Madrid, 28049 Madrid, Spain \\
$^{4}$Institute for Astronomy, University of Edinburgh, Royal Observatory, Edinburgh EH9 3HJ, United Kingdom \\
$^{5}$International Centre for Radio Astronomy Research, University of Western Australia, 35 Stirling Highway, Crawley, Western Australia 6009, Australia \\
$^{6}$National Institute for Astrophysics, Astronomical Observatory of Trieste (INAF-OATs), via Tiepolo 11, 34131 Trieste, Italy \\
$^{7}$Institute for Fundamental Physics of the Universe (IFPU), via Beirut 2, 34014 Trieste, Italy
}}

\date{Accepted XXX. Received YYY; in original form ZZZ}

\pubyear{2021}

\begin{document}
\label{firstpage}
\pagerange{\pageref{firstpage}--\pageref{lastpage}}
\maketitle

\begin{abstract}
We study the connection between morphology and dynamical state of the simulated galaxy clusters in $z\in[0,1.031]$ from {\sc The Three Hundred} Project. We quantify cluster dynamical state using a combination of dynamical indicators from theoretical measures and compare this combined parameter, $\chi$, with the results from morphological classifications. The dynamical state of the cluster sample shows a continuous distribution from dynamically relaxed, more abundant at lower redshift, to hybrid and disturbed. The dynamical state presents a clear dependence on the radius, with internal regions more relaxed than outskirts. The morphology from multi-wavelength mock observation of clusters in X-ray, optical, and Sunyaev–Zel’dovich (SZ) effect images, is quantified by $M$ -- a combination of six parameters for X-ray and SZ maps and the offsets between the optical position of the Brightest Central Galaxy (BCG) and the X-ray/SZ centroids. All the morphological parameters are highly correlated with each other, while they show a moderately strong correlation with the dynamical $\chi$ parameter. The X-ray or SZ peaks are less affected by the dynamical state than centroids, which results in reliable tracers of the cluster density peak. The principal source of contamination in the relaxed cluster fraction, inferred from morphological parameters, is due to dynamically hybrid clusters. Compared to individual parameters, which consider only one aspect of cluster property (e.g. only clumping or asymmetry), the combined morphological and dynamical parameters ($M$ and $\chi$) collect more information and provide a single and more accurate estimation of the cluster dynamical state.
\end{abstract}
\begin{keywords}
galaxies:cluster:general -- Galaxies: clusters: intracluster medium -- methods:numerical
\end{keywords}


\section{Introduction} \label{sec:intro}

Galaxy clusters represent the most massive, gravitationally bound structures in the Universe. The capability to recover a complete description of their gravity potential or their matter distribution is relevant for cosmological studies since their formation and growth are closely related to the underlying cosmological model \citep[e.g.][]{Voit2005,Pratt2019}. Most of the cluster cosmological constraints are based on the mass function, i.e., the number of clusters per mass and redshift bin. However, the cluster total mass is not directly observable but can be inferred through several complementary observational approaches. Some of these are based on certain assumptions on the clusters dynamical state. Even under these hypotheses, the measurement of the mass of clusters is not a simple task because these objects are complex systems made up of several mutually interacting components. Most of the mass in a typical cluster ($M\sim 10^{14}-10^{15}$M$_{\sun}$) is in the form of Dark Matter (DM) that holds together the baryonic components: hundreds of galaxies and the hot X-ray emitting gas or Intra-Cluster Medium (ICM). For a virialized and dynamically ``relaxed'' cluster, the assumption of the hydrostatic equilibrium to describe the gas state might be accurate. However, during merging events or when turbulent motions or compression or non-thermal heating of the ICM dominate, the equilibrium is no more in place and it is not trivial to derive the cluster mass from the radial profiles of the thermodynamical properties of the gas (density, pressure and temperature).

The impact of an `active' dynamical state on the mass reconstruction can be investigated using numerical simulations. Indeed, despite a non-uniform definition of relaxed or disturbed clusters, several authors found similar deviations from hydrostatic equilibrium \citep{Nagai2007, Rasia2012, Henson2016, Biffi2016, Pearce2020} and identify similar causes: turbulence, shock fronts, temperature inhomogeneities in the X-ray-emitting ICM, density inhomogeneities or clumps \citep{Rasia2014, Nelson2014, Biffi2016, Planelles2017, Ansarifard2020}. In this context, the masses of the disturbed clusters are underestimated up to $30\%$, with evidence of mass dependencies \citep{Pearce2020, Gianfagna2021}.

The cluster dynamical state is also linked to other halo properties, such as halo formation time \citep[e.g.][]{Mostoghiu2019, Haggar2020} or halo concentration \citep[e.g.][]{Neto2007}. For these reasons, its determination would be extremely useful for the cosmological use of clusters. Likely, the cluster dynamical state has direct repercussions on the cluster appearance. For this, the cluster morphology has been abundantly studied in the literature, especially using X-ray images \citep[see e.g.][]{Buote1995, Lotz2004, Rasia2013, Parekh2015, Lovisari2017, Lopes2018, Chong2018, Cialone2018}.

This work continues this series of investigations by extending the analysis to unprecedented statistics of massive clusters. We use the galaxy cluster catalogues from {\sc The Three Hundred} Project\footnote{\url{https://the300-project.org}}: a set of 324 cluster-centric regions of 15 $h^{-1}$Mpc radius simulated with hydrodynamics which includes radiative physics and different sub-grid models to describe the stellar and black-holes populations. For each cluster, we produce and analyse optical, X-ray and $y$ maps, where the $y$ maps are the distribution of the Comptonization parameter $y$, related to the thermal Sunyaev-Zel'dovich effect \citep[or tSZ,][]{Sunyaev1972,Sunyaev1980} and observed in the microwave band as a distortion of the Cosmic Microwave Background (CMB). Our specific goal is to determine how to best use the morphological information derived from these maps to efficiently describe the true dynamical state of a cluster, which in this work is parametrized by theoretical indicators computed directly from the 3D information of the simulated clusters.

This paper is structured as follows: in Sec.~\ref{sec:data_an} we present some details of the simulations, the cluster catalogues, and the synthetic maps. In Sections \ref{sec:DS} and \ref{sec:2D} we introduce the dynamical and morphological indicators used in our analysis. Our results are discussed in Sec.~\ref{sec:results} and our findings are summarised in Sec.~\ref{sec:conclusion}.

\section{Dataset} \label{sec:data_an}

\subsection{The Three Hundred cluster catalogue} \label{ssec:sample}

The numerical cluster samples studied in this work belong to {\sc The Three Hundred} Project, introduced in \citet{300} and used in \citet{Wang2018, Mostoghiu2019, Arthur2019, Haggar2020,Kuchner2020}. This consists of a series of zoomed hydrodynamic simulations of 324 cluster regions extracted from \textsc{MDPL2}, \textit{The MultiDark Planck 2} simulation \citep{Klypin2016}, a 1$h^{-1}$Gpc DM-only simulation with a cosmology consistent with \citet{pl15}. The clusters were initially selected from \textsc{MDPL2} simulation by their virial\footnote{In this paper, we indicate with $R_{\Delta}$ the radius of the sphere whose density is $\Delta$ times the critical density of the Universe at that redshift $\rho(R_\Delta)=\Delta\rho_{cr}(z)$. We specifically use overdensities equal to $\Delta=500$, $200$, and vir, where $\Delta_{vir}$ corresponds roughly to 98 for the assumed cosmological model.} halo mass ($M_{vir}>8\times10^{14}h^{-1}$M$_{\sun}$ at $z=0$), which is identified by the Rockstar halo finder \citep{Behroozi2012}. This results in the most massive 324 clusters which were used to regenerate the zoomed-in initial conditions for the hydrodynamic runs. As shown in the appendix of \citet{300}, most of them are still the most massive clusters at $M_{200}$ and $M_{500}$ but with a slightly lower value for the mass completeness thresholds: $M_{200} > 6.4\times10^{14}$ and $M_{500} > 4.6\times10^{14}$. The Lagrangian areas of these spherical regions were computed from a low-resolution version of the MDPL2 and initial conditions were produced using the \textsc{ginnungagap} code \citep{300}, with multiple levels of mass refinements, keeping the original mass resolution of the MDPL2 simulation for the particles within the Lagrangian region and spawning one gas particle per DM particle. Accordingly to the Planck estimate of the cosmic baryon fraction, the gas particles in the highest-resolution volume have an initial mass equal to $2.36\times10^8h^{-1}M_{\sun}$ while the mass of the DM particles is $1.27\times10^9h^{-1}M_{\sun}$. In order to reduce the computational costs of the simulations, the mass resolution of Dark Matter outside the Lagrangian region has then been degraded in such a way as to preserve the same tidal field.

Within {\sc The Three Hundred} Project, the same 324 Lagrangian regions are re-simulated with different codes, however, for the specific analysis presented here, we focus only on the catalogues extracted from the \textsc{gadget-x} hydrodynamical simulations \citep{Murante2010,Rasia2015}. This code is a modified version of \textsc{gadget3} Tree-PM code and includes an improved SPH scheme with Wendland interpolating C4 kernel, artificial thermal diffusion and time-dependent viscosity. Other main features of these runs are gas cooling with metal contributions, star formation with chemical enrichment and feedback from stars in the asymptotic giant branch, supernovae, and active galactic nuclei. For a more detailed description of {\sc The Three Hundred} Project, see the recent works based upon these simulations of \citet{ Li2020, Knebe2020, Rost2021, Mostoghiu2021, Kuchner2021}.

During the simulation production phase, we store the data for 128 different snapshots in the redshift range between $z=17$ and $z=0$. In this work we analyse clusters coming from 10 selected redshifts: $0$, $0.116$, $0.193$, $0.304$, $0.457$, $0.557$, $0.663$, $0.817$, $0.900$ and $1.031$. This choice has been made to study the redshift evolution of both the morphological parameters and the dynamical state indicators. The partial redshift overlap with \citet[][hereafter \citetalias{Cialone2018}]{Cialone2018} allows us to compare the results with those from the \textsc{MUSIC} simulation. For each region of the simulation, halos and sub-halos are identified with the Amiga Halo Finder, AHF\footnote{\url{http://popia.ft.uam.es/AHF}} \citep{AHF}, whenever the structure has at least 20 particles. From the output of AHF, we select for the analysis the most massive central clusters at each redshift, for a total of 3240 objects. The mass range of the galaxy clusters is $M_{500}=(0.15$-$17.58)\times10^{14}h^{-1}$M$_{\sun}$ (median $2.79\times10^{14}h^{-1}$M$_{\sun}$).

\subsection{Mock optical, X-ray and SZ maps} \label{ssec:maps}

We generate three maps per cluster reproducing optical, X-ray and millimetre observations. The last category is aimed to mimic maps from the tSZ effect. They are produced considering a spherical region of radius $1.4R_{200}$, centred on the projected position of the theoretical cluster centre defined here as the maximum of the density. In order to mimic observation maps, clusters at $z=0$ are replaced at $z=0.05$ for the three maps with different angular resolutions. Only the projection along the z-direction is used in this paper. However, we note here that the other projections give similar results. All synthetic maps are produced without including the contribution of other sky contaminants or instrumental noise. Finally, the resolution of each map is specified based on the target observation. \\

The {\bf optical maps} of the clusters reproduce the optical r band of the Sloan Digital Sky Survey (SDSS), with the same angular resolution of $0.396\arcsec$ per pixel. The main sources in this band are galaxies whose stellar luminosities are derived applying a stellar population synthesis code described in \citet{Cui2011, Cui2014, Cui2016}. Each star particle in simulation is treated as a simple stellar population with a Chabrier initial mass function \citep{Chabrier2003}, which is also adopted in \textsc{gadget-x}. The spectrum from each star particle is thus produced by interpolating the stellar evolution library of \citet{Bruzual2003} with its metallicity and age. Then the spectra of the star particles within the same pixel are sum up to convolve with the SDSS response file to produce the r-band luminosity which is in units of erg s$^{-1}$ cm$^{-2}$. \\

In {\bf X-ray band}, galaxy clusters are strong and extended sources. The X-ray emission is due to the process of thermal bremsstrahlung in which hot electrons are scattered by ions in the ICM. The surface brightness $\Sigma_X$ along the line of sight can be written as:
\begin{equation}
\label{eq:bremm}
\Sigma_X(\nu)=\frac{1}{4\pi(1+z)^3}\int n_e n_i\Lambda_X(T,Z,\nu)dl,
\end{equation}
where $n_e$, $n_i$ are the electrons and ions number densities and $\Lambda_X(T,Z,\nu)$ is the cooling function which depends on the frequency $\nu$, the metal abundances $Z$ and the temperature of the plasma $T$. X-ray images are produced using \textsc{pyXSIM code} \citep{pyXSIM, pyXSIM2} based on the PHOX algorithm \citep{phox}. We adopt the APEC model from AtomDB\footnote{\url{http://www.atomdb.org/index.php}} as the thermal spectral model in \textsc{pyXSIM} for generating photons. We further include the Tuebingen-Boulder \citep{Wilms2000} absorption model with the neutral hydrogen column density in units of $0.1\times 10^{22} {\rm atoms\ cm}^{-2}$ for the foreground galactic absorption. The X-ray maps are in terms of number counts of detected photons with 10ks exposure time and their spectral band is $0.1$-$15 keV$. We use the responses associated with the WFI instrument which will be on-board the Athena satellite \citep{Athena}.\\

The {\bf SZ effect} is originated through inverse Compton scattering of CMB photons with ICM hot electrons. The distortion is caused both by the random thermal motion of electrons (thermal SZ effect) and by the overall bulk motion of the cluster with respect to the Hubble flow (kinematic SZ effect). Cluster maps in microwave band are dominated by the thermal SZ, since for the expected velocities of galaxy clusters (few hundred km s$^{-1}$), and typical cluster temperatures (few keV) the kinematic contribution is about 10 per cent of the thermal one \citep{Birkinshaw1999, Carlstrom2002}. Therefore in our analysis, we study only the thermal SZ maps that could be described in terms of the 2D distribution of the dimensionless Comptonization parameter $y$. It is defined as:
\begin{equation}
\label{eq:y}
y=\int n_e\sigma_T\frac{kT_e}{m_ec^2}dl,
\end{equation}
where $m_e$ and $T_e$ are the electron mass and temperature, $\sigma_T$ is the Thomson cross section, $c$ the speed of light, $k$ the Boltzmann constant and $dl$ is the line of sight length. Operationally, we compute a discretised version of Eq.~\eqref{eq:y} for which we assume that $dV=dAdl$ and $dA$ is the pixel area \citep{MUSIC, 300}:
\begin{equation}
\label{eq:y_disc}
y=\frac{\sigma_T k}{m_ec^2dA}\sum_i T_i N_{e,i} W(r,h_i),
\end{equation}
where $N_{e,i}$ is the number of electrons, $h_i$ the SPH smoothing length and $W(r,h_i)$ the SPH smoothing kernel used in the simulation. The $y$ maps are produced by the \textsc{pyMSZ} code\footnote{\url{https://github.com/weiguangcui/pymsz}}, which can also generate the kinematic SZ effect maps simultaneously \citep[see][for an application to the MUSIC simulation]{Baldi2018}. By passing the cluster centre and radius, the package will load the simulation snapshot (it supports different snapshot formats) for all necessary information for calculation. It will output the y-map in fits file with the given projection direction, angular resolution and the redshift where the cluster locates.

Both ICM maps have a fixed spatial comoving resolution of $10$ kpc pixel$^{-1}$. Notice that the X-ray and $y$ maps give complementary information about the cluster structure. SZ effect data are more effective in describing the cluster outskirts compared to X-ray images since the $y$ signal is roughly linearly dependent on the electron density while the X-ray emission is instead proportional to density square.

The following analysis based on the maps considers the map centroids as the centre of reference instead of the theoretical cluster centre to not bias our results by a priori knowledge of the true cluster centre. The centroids of the X-ray and $y$ maps are calculated considering the emission of all pixels within a circle of radius equal to $R_{500}$, centred on the theoretical cluster centre.

All maps are used to extract the morphological parameters described in Sec.~\ref{ssec:2D_morp}, while in the next Section we introduce the indicators of the dynamical state computed using the 3D information.

\section{Dynamical state indicators} \label{sec:DS}

In the case of hydrodynamical simulations, all the physical properties of each particle are known. Therefore for a given object, it is possible to estimate all the physical quantities in interest, such as density, gravitational potential, pressure, mass, etc. The theoretical indicators of the dynamical state applied to simulations use this advantage and thus refer to quantities computed in 3D that would be unreachable from an observational analysis. \citet{Barnes2017} and \citet{Pearce2020} consider, for example, the ratio between the kinetic and thermal energy of the particles inside the halo to estimate the dynamical state of the clusters, while the dimensionless measure of the Dark Matter halo rotation (the spin parameter $\lambda$) is used in \citet{Maccio2007, Klypin2011}.

Throughout this paper, we use five indicators of the cluster dynamical state: $(i)$ the mass fraction of all sub-halo in the cluster, $f_{s}$, $(ii)$ the ratio between the masses of the most massive substructure and the cluster, $f_{s,mm}$, $(iii)$ the offset between the cluster centre and the centre of mass, $\Delta_r$, $(iv)$ the ratio between thermal and potential energy, $\eta$, and $(v)$ the relaxation parameter $\chi$ \citep{Haggar2020}. In the following, we will describe each of them in more detail, but not before underlining that in the literature there are many applications of these parameters for the relaxation definition \citep[see][and references therein]{Neto2007, Ludlow2012, Ludlow2014, Meneghetti2014, Henson2016, Planelles2017}.
\\

By identifying with AHF all the sub-halos present inside a spherical region of a cluster with radius $R_{\rm ap}$, the total sub-halo mass fraction $f_s$ is defined as the ratio between the sum of all the sub-halo masses and the cluster mass within such volume, $M_{\rm ap}$:
\begin{equation}
	f_{s}=\dfrac{\sum_i M_{i}}{M_{\rm ap}}.
	\label{eq:fs_t}
\end{equation}

The other mass fraction indicator, $f_{s,mm}$, is built considering only the contribution of the most massive substructure in the cluster:
\begin{equation}
  f_{s,mm}=M_{mm}/M_{\rm ap}.
 \label{eq:fs_m}
\end{equation}

The virial ratio $\eta$ is based on the virial theorem and it is defined as:
\begin{equation}
	\eta=\frac{2T-E_s}{|W|},
	\label{eq:eta}
\end{equation}
where $T$ is the total kinetic energy, $E_s$ is the surface pressure energy from both collisionless and gas particles, and $W$ is the total potential energy \citep[see][for details or applications]{Klypin2016, Cui2017, Reju2019}.

The offset of the centre of mass $\Delta_r$ is widely used in the literature \citep[e.g.][]{Maccio2007, Maccio2008, Duffy2008, Sembolini2014}. It is quantified as:
\begin{equation}
	\Delta_r=\dfrac{|\mathbf{r_{cm}}-\mathbf{r_c}|}{R_{\rm ap}},
	\label{eq:3D_off}
\end{equation}
where $\mathbf{r_{cm}}$ is the centre-of-mass position of the cluster and $\mathbf{r_{c}}$ is the theoretical centre of the cluster which we identify as the position of the highest density peak. 

Finally, in order to describe the degree of relaxation \citet{Haggar2020} proposed to use the inverse square root of the normalised quadratic mean of various indicators, generically indicated as $x_i$:
\begin{equation}
    \chi=\left[\frac{\sum_i \left(\frac{x_i}{x_{0,i}}\right)^2}{N}\right]^{-1/2},
    \label{eq:chi}
\end{equation}
where $x_{0,i}$ are the classification thresholds used to distinguish between relaxed and disturbed clusters. 

Unfortunately, in literature, there is not a unique selection of these thresholds and also of the set of 3D dynamical indicators ($x_i$) that are the most suitable to segregate among relaxed and disturbed clusters \citep[see also][]{Cui2017}. The variety of choices made by different authors is partially justified from the fact that different kinds of simulations were involved (e.g. DM versus hydrodynamical runs with different treatments for the baryon physics) or because the dynamical state indicators were extracted from different volumes such as those within $R_{\rm vir}$ or $R_{200}$ or $R_{500}$. In fact, by including the most external regions, the cluster will be less virialized, which could be caused by the inclusion of more substructures that are still in the process of merging into the cluster. Studying these dependencies is one of the goals of this paper.

In addition to the usage of a continuous parameter, such as the combined $\chi$ parameter defined above, we also classify the clusters in three separate classes called `relaxed', `hybrid', and `disturbed'. These classes are defined by using the parameters $f_s$ (Eq.~\eqref{eq:fs_t}) and $\Delta_r$ (Eq.~\eqref{eq:3D_off}). Specifically, we defined relaxed (disturbed) all objects for which the two conditions $f_s<0.1$ and $\Delta_r<0.1$ ($f_s>0.1$ and $\Delta_r>0.1$) are simultaneously verified. The hybrid class includes all other clusters, i.e., those for which the two inequalities have different signs. This class-based division will be useful to compare with other works present in the literature.

In this work, we study the dynamical state of {\sc the Three Hundred} clusters in Sec.~\ref{ssec:3D_ds} and we compare the results of different relaxation criteria on {\sc The Three Hundred} sample in Sec.~\ref{ssec:dyn_comp}. In particular, we compare the result of the relaxation criteria used in \citet{300} and \citetalias{Cialone2018} with our findings. We select and tune the best morphological parameters among those that better segregate relaxed from disturbed clusters by using as prior our knowledge on the systems’ dynamical state as measured from the 3D dynamical indicators. The procedure will be described in Sec.~\ref{sec:2D} and applied in Sec.~\ref{ssec:res_Xy}.

\section{Morphological indicators} \label{sec:2D}

Historically, the morphology of clusters has been studied using several parameters applied to the different multi-wavelength maps \citep[e. g.][with references therein]{Okabe2010, Meneghetti2014, Lovisari2017, Bartalucci2019, Cao2021}. Most of the ICM morphological indicators have been originally introduced for X-ray cluster maps \citep{ Santos2008, Nurgaliev2013, Mantz2015} to detect the presence of substructures \citep{Mohr1993, Buote1995, Poole2006, Jeltema2008} and were borrowed and adapted from optical studies on the galaxy morphology \citep{Rasia2013}, or even from optical analysis as the application of Zernike polynomials to cluster maps \citep{Capalbo2020}. The cluster dynamical state can also be inferred from some optical substructure estimators \citep{Pinkney1996, Roberts2018}, based on galaxies properties such as local deviations from global mean and dispersion of radial velocities, magnitude difference between the Brightest Central Galaxy (BCG) and the second brightest galaxy \citep[e.g.][and reference therein]{Lavoie2016, Lopes2018}, and offsets between the BCG and the X-ray peak or X-ray centroid \citep{Sanderson2009, Mann2012, Mahdavi2013, Cui2016, Rossetti2016, Lopes2018, Zenteno2020}.

In this paper, we apply six ICM morphological indicators on both mock X-ray and $y$ maps, plus a combination of them. The combination of various parameters into one is a strategy already used in literature \citep{Rasia2013} since each parameter highlights only a particular aspect of a typical disturbed system and, at times, the efficacy of one parameter in describing the cluster dynamical status depends on the chosen line of sight as projections might influence the result \citepalias{Cialone2018}. Together with this set of parameters based on the ICM appearance, we also study parameters based on the offsets between BCG and X-ray and $y$ peaks or centroids positions. 

The definitions of all these parameters are described in Sections \ref{ssec:2D_morp} and \ref{ssec:2D_offset}, while in Sec.~\ref{ssec:2D_morp_eff} the diagnostic ability of morphological parameters is studied by using the Kolmogorov-Smirnov (KS) two-tail test, and the analysis of the Receiver Operating Characteristic (ROC) curve \citep[see e.g.][for a more detailed introduction of ROC diagnostic test]{Swets1988, Fawcett2006}. Finally, the segregation ability of the morphological parameters is tested comparing them with the 3D dynamical indicator $\chi$ in Sec.~\ref{sec:results}.

\subsection{ICM morphological indicators} \label{ssec:2D_morp}

The morphological indicators for X-ray and $y$ maps used in this work are the same as described in \citetalias{Cialone2018} for the MUSIC simulation:
\begin{description}
\item[$A$,] Asymmetry \citep{Schade1995, Okabe2010, Zhang2010} is a normalised difference between the original map and a rotated one. For our analysis, we analyse 4 different rotations of the maps ($90\degr$, $180\degr$ and the flipped images along the main axes) and then we consider, for each cluster, the rotation which maximises $A$;
\item[$c$,] Light Concentration Ratio \citep{Santos2008} is the ratio of the surface brightness, computed inside two concentric apertures;
\item[$w$,] Centroid Shift \citep{Mohr1993, Poole2006, OHara2006, Bohringer2010} is the average of the shifts of the centroids obtained from various concentric circles with increasing radius;
\item[$P$,] Power Ratio \citep{Buote1995} is based on a multipole decomposition applied to the maps of the ICM which are thought to represent the projected mass distribution;
\item[$G$,] Gaussian Fit \citepalias{Cialone2018} is the ratio of the two standard deviations of a 2D Gaussian fit to the X-ray and $y$ maps;
\item[$S$,] Strip \citepalias{Cialone2018} is defined as a normalized difference of $N$ light profiles, passing through the centroid. Following \citetalias{Cialone2018}, we compare 4 strips inclined by $45\degr$ to each other.
\end{description}
As the definition of the 3D dynamical indicators, the morphological indicators depend on the aperture, $R_{\rm ap}$, used to estimate them. To determine which aperture is the most efficient in separating the clusters, we employ the same procedure illustrated in \citetalias{Cialone2018}. For each aperture, we create two distributions of the morphological parameters relative to the clusters of both the relaxed and disturbed classes introduced at the end of the last Section. With these two distributions as input, we compute the KS test and consider as the best aperture the one that corresponds to the minimum of the median of all KS-p values, over the entire redshift range. The results of this tuning are shown in Sec.~\ref{ssec:res_Xy}. All these tuned parameters $V_i$ are then collected in the combined parameter $M$, defined as in \citetalias{Cialone2018}:
\begin{equation}
M=\frac{1}{\sum_i W_i}\left(\sum_i W_i \dfrac{\log_{10}(V_i^{\alpha_i})-<\log_{10}(V_i^{\alpha_i})>}{\sigma_{\log_{10}(V_i^{\alpha_i})}}\right),
\label{eq:M}
\end{equation}
where $\alpha_i=\pm1$, depending on how the $i$-th parameter is related to the dynamical state. $M$ parameter represents a weighted average of the standardised indicators described above, to enhance and have a single parameter to characterise the morphology. The logarithm of the minimum KS-p median value over the entire redshift range is used as a weight, $W_i$, in $M$ definition (Eq.\eqref{eq:M}), for each parameter:
\begin{equation}
    W_i=\left|\log_{10}\left(min\left\{\widetilde{KS}\textnormal{-}p_{i,z}\right\}_{R_{ap}}\right)\right|.
\label{eq:Wi}
\end{equation}

\subsection{Offset morphological indicators} \label{ssec:2D_offset}

\begin{figure}
	\includegraphics[width=\columnwidth]{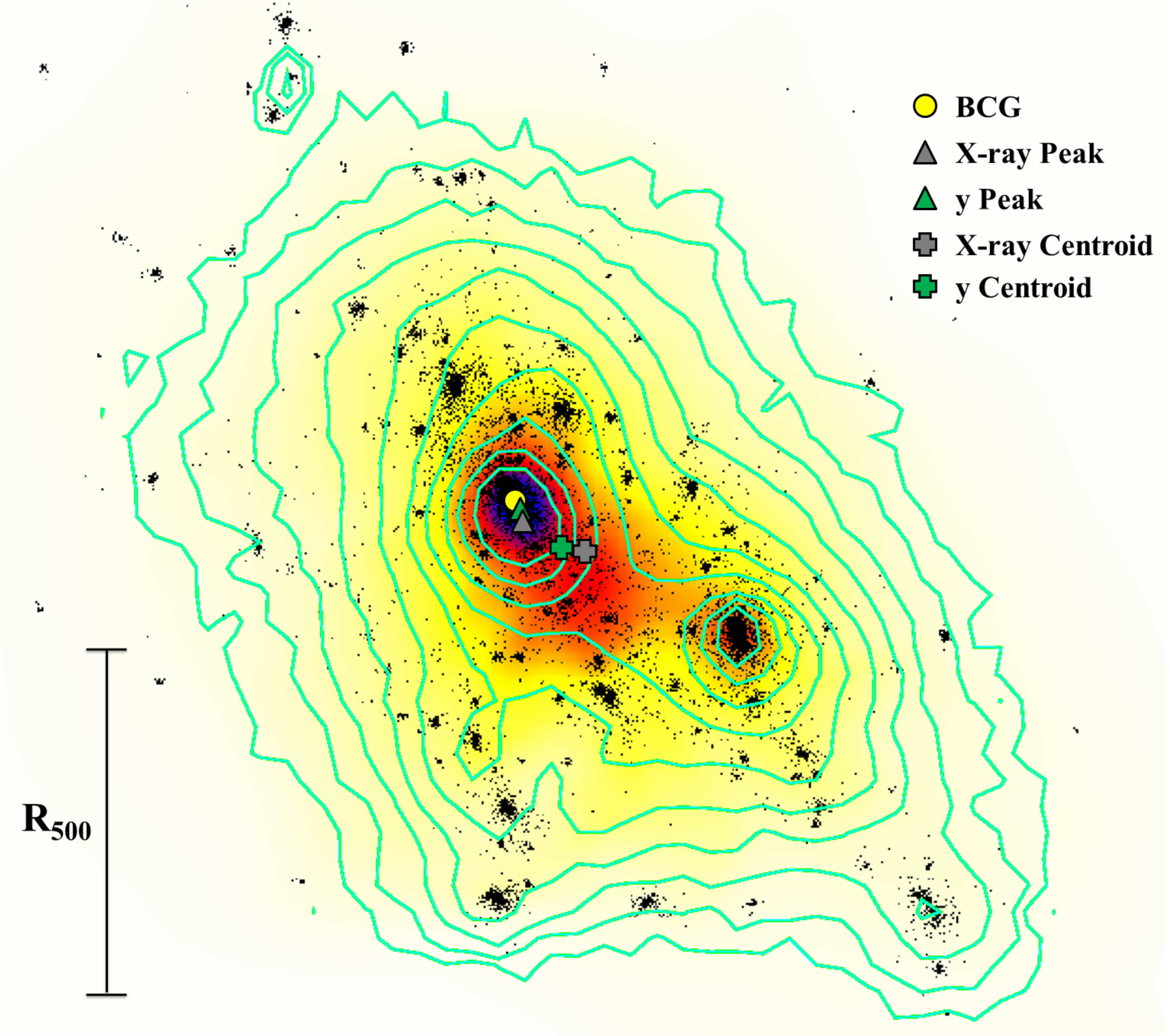}
    \caption{An illustrative example of a combination of the mock multi-wavelength maps described in Sec.~\ref{ssec:maps}, for the central galaxy cluster at $z=0.46$, in region 85 of the simulation. The $y$ colourmap and the contours of the X-ray photon counts, in log scale, are superimposed on the optical SDSS r band. Isocontour levels are logarithmic equispaced by a factor $\log2$. The symbols in the figure mark the position of the relevant cluster centres for the offset parameters, described in Sec.~\ref{ssec:2D_offset}. The position of the BCG is marked with a yellow dot, the X-ray and $y$ peaks with a grey and green triangle, and the centroids with crosses, respectively, of the same colours. The $R_{500}$ radius of the cluster is shown in the left corner of the figure.}
    \label{fig:all}
\end{figure}

On top of the ICM-based morphological indicators, we use also the offset between the BCG position and both the centroids and the peaks of the X-ray and $y$ maps \citep{Lavoie2016, Lopes2018}. As an example, we visualise all relevant positions in Fig.~\ref{fig:all}. In general, the BCG position is expected to trace the position of the matter density peak inside clusters \citep{Cui2016}, as postulated in the "Central Galaxy Paradigm" \citep{Tremaine1990, Postman1995, Lin2004, Lopes2018}. We discuss the validity of the Central Galaxy Paradigm and the efficiency of these offset parameters in Sec.~\ref{ssec:2D_off_res}.

\subsection{Methods to estimate the efficiency of the morphological parameters} \label{ssec:2D_morp_eff}

In this work, we study the performance of our morphological classifiers using two different tests: the Kolmogorov-Smirnov test and the analysis of ROC curves to which we associate and study several diagnostic parameters. The KS test is a statistical non-parametric test that determines whether two samples are representative of the same distribution by comparing their cumulative distribution function. This test returns the maximum deviation between the two curves and a parameter referred to as $p$ value which provides the significant level of the result. A small $p$ value implies that the two distributions are different.

As explained in Sec.~\ref{ssec:2D_morp}, the KS test is used to retrieve the best aperture to calculate the six morphological parameters. In particular, using the dynamical state classification as a prior, we can compare the relaxed and disturbed distributions of the morphological parameters estimated in different aperture with the KS test. This process is repeated for all redshifts. Then for each morphological parameter, we consider the median of the KS-p values as a reference to determine the best overall aperture. These medians are used also to compute the combined parameters, as in Eq.~\eqref{eq:Wi}.

Generally, for the classification of clusters in observations, continuous morphological parameters $V$ are applied to divide them into sub-samples. To do that, a threshold $V_T$ is selected on those parameters above (below) at which the clusters are morphological regular (disturbed). This classification will reflect the dynamical state according to the efficiency of the parameters, which can be described in terms of false and true detections. If a cluster is dynamically relaxed but does not satisfy the morphological threshold, we can define this case as a false negative ($FN$), wrong classification. Classifying instead a disturbed cluster as regular we will have a false positive ($FP$) case. Vice versa, the proper selections are defined as true positive ($TP$, the morphologically and dynamically relaxed) or true negative ($TN$, the morphologically and dynamically disturbed) objects. All these outcomes are generally collected together in the contingency (or confusion) matrix. Several evaluation metrics can be defined from these four classes, as the completeness ($C$), the purity ($p$) \citep{Rasia2013}, or the Matthews correlation coefficient ($MCC$). The selection of the threshold $V_T$ is crucial for the classification, since changing this value we will modify the result of the classification and the diagnostic power of the used classifier. To characterise that dependence for our morphological parameters, we study the ROC curves associated with the dynamical state described by the three classes defined in Sec.~\ref{sec:2D}. 

\paragraph*{Completeness.} \label{par:Completeness}
The completeness quantifies how many correct identifications are performed in the test and it is defined as the true positive rate, $TPR$, the number of correct classifications divided by the total number of relaxed clusters:
\begin{equation}
    C=TPR=\frac{TP}{TP+FN}.
    \label{eq:completeness}
\end{equation}

\paragraph*{Purity} \label{par:Punrity}
The purity describes the presence of contaminants in the selected sub-sample of only regular clusters, and it is defined as:
\begin{equation}
    p=\frac{TP}{TP+FP}.
    \label{eq:purity}
\end{equation}

\paragraph*{Matthews correlation.} \label{par:MCC}
The $MCC$, equivalent to the Pearson $\phi$ coefficient, is defined considering all the terms of the confusion matrix, taking care of unbalanced samples distribution:
\begin{equation}
    MCC=\frac{TP \times TN - FP \times FN}{\sqrt{(TP+FP)(TP+FN)(TN+FP)(TN+FN)}}.
    \label{eq:MCC}
\end{equation}

\paragraph*{ROC curves.} \label{par:ROC}
Leaving the threshold $V_T$ to vary, the ROC curve is defined as the graph of $TPR$ (the completeness $C$) against the false positive rate $FPR$, the number of disturbed clusters incorrectly recognised as relaxed, in terms of the total number of disturbed objects:
\begin{equation}
    FPR=\frac{FP}{TN+FP}.
    \label{eq:FPR}
\end{equation}
The ROC curve is a powerful graphical test: in the case of a perfect classifier, the associated ROC curve will be described in the $TPR$-$FPR$ plane by a unit step function. On the contrary, an indicator that has an equal probability to recognise a cluster as relaxed or disturbed is instead described in the same plane by the identity line. From the properties of this curve, several summary statistics for the diagnostic power are commonly drawn, such as the area under the curve ($AUC$), with $AUC=0.5$ associated with random guess and $AUC=1$ to the perfect case, or the Youden's $J$ statistics. $J$ is defined as: 
\begin{equation}
    J=TPR+TNR-1,
    \label{eq:Jstat}
\end{equation}
where $TNR$ is the true negative rate (the number of clusters that are correctly recognised as non-relaxed over the total number of non-relaxed clusters). It represents, graphically, the ROC height above the random guess line.

\paragraph*{Probability.} \label{par:Prob}
Another simple way to estimate the diagnostic ability of the parameters and the contamination of non-relaxed classes is to define a probability, $P$, to count in our sample a relaxed, hybrid or disturbed cluster for a given value of the classifier $V$. A simple merit function to quantify this can be defined as:
\begin{equation}
	P_z^{r,h,d}(V)=\dfrac{N_z^{r,h,d}(V)}{N_z^{r}(V)+N_z^{h}(V)+N_z^{d}(V)},
	\label{eq:IG_rd}
\end{equation}
where $N_z^{r,h,d}$ is the number of relaxed (subscript r), hybrid (h) or disturbed (d), objects that have a certain value $V$ for redshift $z$. The purity $p$ corresponds to the integral of $P$: $p=\int P dV$. \\

In this work, we use the ROC curve, $AUC$, $J$, $MCC$, $C$, $p$ and $P$ to study the efficiency and the purity of sub-samples when a threshold is applied to morphological parameters. In particular, we use $MCC$ and $J$ to infer a suitable and not arbitrary threshold ($V_T$) on morphological parameters to divide relaxed objects from the other cases. In fact, $J$ and $MCC$ can be used as a score for the performance of the test: their (absolute) value ranges from 1 through 0, depending on whether the test is able or not to discriminate between the two classes. Considering that the performance of the test changes if the discrimination threshold is varied, we can choose as threshold the one that maximises these two evaluation metrics. A detailed discussion of the consistency of relaxed sub-samples inferred with different criteria is beyond the goal of this paper. However, we still compare the fraction of relaxed clusters available in the literature \citep[see also][and references therein]{Rasia2013, Mantz2015, Rossetti2016, Lovisari2017, Cao2021} with our findings in Sec.~\ref{ssec:2D_obs_comp}.

\section{Results} \label{sec:results}

In this Section, we discuss the dynamical state of clusters and the efficiency of the morphological parameters described in Sections \ref{ssec:2D_morp} and \ref{ssec:2D_offset}. Then in Sec.~\ref{ssec:2D_obs_comp}, we compare our morphological results with other clusters samples, available in the literature.

\subsection{Dynamical state of The Three Hundred Galaxy Clusters} \label{ssec:3D_ds}

\begin{figure}
	\includegraphics[width=\columnwidth]{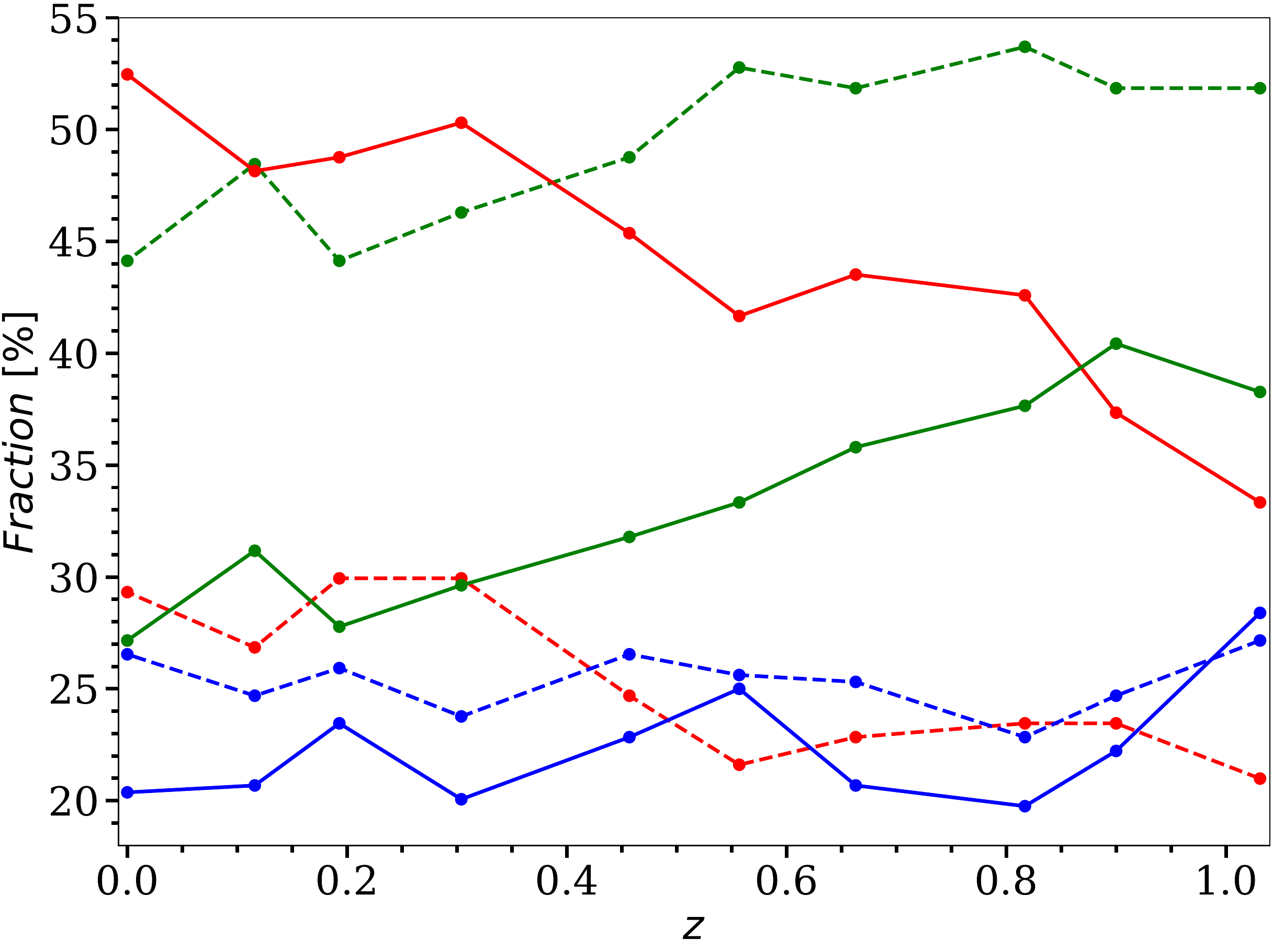}
    \caption{Percentage of relaxed (red lines), hybrid (green), disturbed (blue) clusters along the redshift, using the dynamical state indicators and the relaxation criterion discussed in Sec.~\ref{sec:DS}. The solid and dashed lines show the results using a volume radius of $R_{500}$ or $R_{200}$, respectively.}
    \label{fig:3D_dyn}
\end{figure}

In Fig.~\ref{fig:3D_dyn} we show the percentage of relaxed, hybrid and disturbed classes, defined in Sec.~\ref{sec:DS}, as a function of redshift. Dashed and solid lines refer to measurements done within $R_{200}$ and $R_{500}$, respectively. The relaxed and hybrid populations show a redshift evolution with reverse trends: the fraction of relaxed clusters decreases while the hybrid fraction increases from $z=0$ to $z=1$. At the same time, the disturbed class remains almost constant. This redshift evolution is expected since clusters start to relax at about $z\sim1$ but the majority reaches a virialization status only by $z=0$ \citep{Muldrew2015}. The hierarchical cluster evolution can also explain the quite different percentage of objects defined as relaxed within $R_{500}$ (above 50 per cent at $z=0$) and within $R_{200}$ (30 per cent at $z=0$). The strong decrease associated with the largest volume suggests that several substructures are present in the cluster outskirts and they have not reached a relaxation status yet. Although, since the disturbed class does not dramatically change, the cluster outskirts affect only one of the two parameters entering into the relaxation definition (either $f_{s}$ or $\Delta_r$). As already noted, we recall that the exact value of the relaxed cluster fraction depends not only on the volume considered but also on the chosen threshold ($x_0$) as we will explore in the next Section and in Fig.~\ref{fig:rel_com}.

\begin{figure}
	\includegraphics[width=\columnwidth]{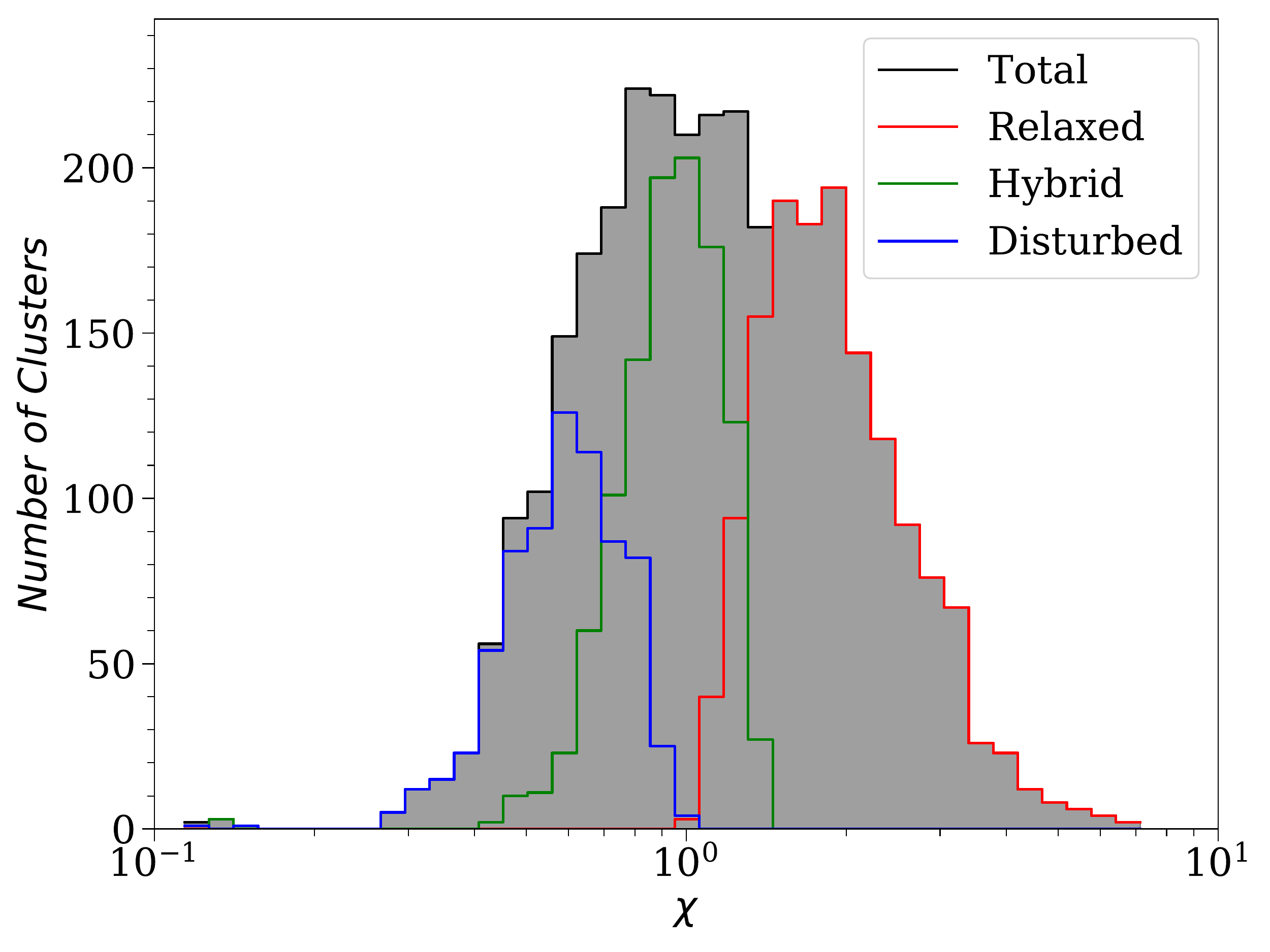}
    \caption{Distribution of the combined $\chi$ indicator, defined in Sec.~\ref{sec:DS} and using in its definition the dynamical indicators calculated inside $R_{500}$. All the 3240 studied galaxy clusters are considered in the distribution, with blue, green and red curves corresponding to the disturbed, hybrid and relaxed classes.}
    \label{fig:chi}
\end{figure}

For the study of the connection between the dynamical state and morphology of clusters, we decide to limit our analysis only to regions inside $R_{500}$. This choice was made to study the same region commonly achievable in observations and, therefore, we will consider the dynamical state defined within this aperture. In Fig.~\ref{fig:chi} we show the distribution of the continuous $\chi$ dynamical indicator obtained from the same parameters ($x_i=(f_s,\Delta)$) and thresholds ($x_{0,i}=0.1$ for both) that we use for the dynamical classification scheme. The distribution is drawn including all clusters at all redshifts. Over-plotted we also show the distributions of the relaxed, hybrid, and disturbed classes (see Sec.~\ref{sec:DS}) in red, green, and blue, respectively. By definition, relaxed (disturbed) systems have $\chi$ greater (lower) than 1. The hybrid systems, instead, occupy the region between the two extreme population.

\subsection{Impact of different criteria on dynamical state} \label{ssec:dyn_comp}

In this Section, we want to compare our findings with previous results in the literature. For this, we adopt the same criteria used in \citetalias{Cialone2018} and \citet{300} and, as before, we compute the dynamical indicators within both $R_{500}$ and $R_{200}$. In \citetalias{Cialone2018} the dynamical state of the MUSIC galaxy clusters is studied within $R_{vir}$, using $\Delta_r$ and $f_{s,mm}$. Relaxed clusters are those who have at the same time $\Delta_r$ and $f_{s,mm}$ less than $0.1$. Instead, in \citet{300} {\sc The Three Hundred} relaxed clusters are defined by adopting more stringent criteria since all the following conditions, measured within $R_{200}$, needed to be simultaneously satisfied: $|1-\eta|<0.15$, $\Delta_r<0.04$ and $f_s<0.1$.

\begin{figure}
	\includegraphics[width=\columnwidth]{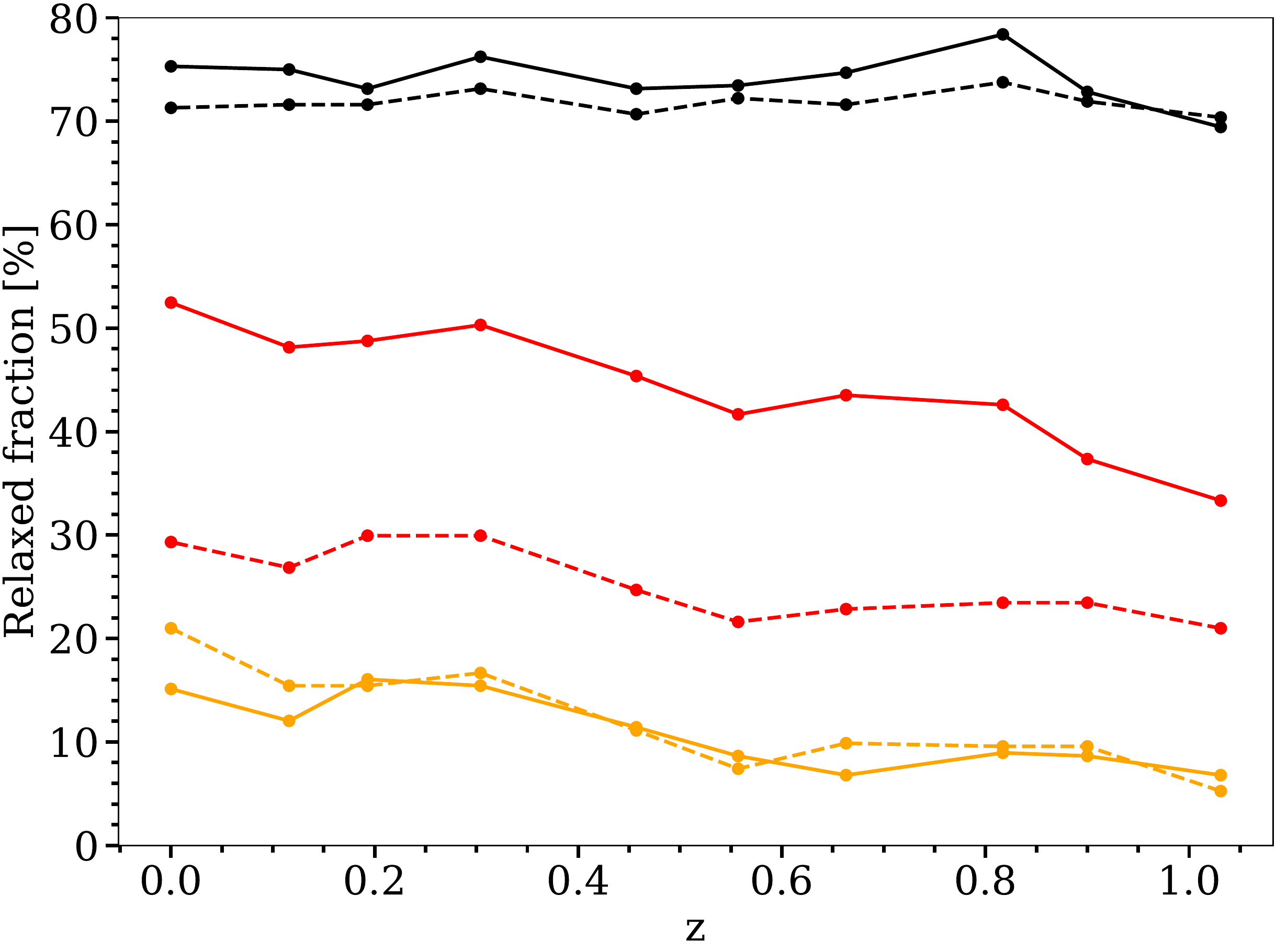}
    \caption{Percentage of relaxed clusters applying the relaxation criteria adopted in this work (Sec.~\ref{sec:DS}, red lines), in \citet{300} (orange) and \citetalias{Cialone2018} (black), in the examined 10 redshift snapshots of the simulation. Dashed and solid lines show the results using respectively $R_{200}$ or $R_{500}$.}
    \label{fig:rel_com}
\end{figure}

The fractions of {\sc The Three Hundred} relaxed clusters adopting these criteria, calculated for $R_{200}$ (solid lines) and $R_{500}$ (dashed lines) are shown in Fig.~\ref{fig:rel_com}. Applying the \citetalias{Cialone2018} criteria to our samples, we recover one of the results highlighted in that paper: the fraction of relaxed clusters is constant in the redshift range considered. Note here that it is more extended ($0<z<1$) than in \citetalias{Cialone2018} ($0.4<z<0.8$). Furthermore, the relaxed fraction remains almost constant both considering $R_{500}$ or $R_{200}$. It is, however, striking the difference in the number of relaxed clusters: up to 70\% here (black lines) and close to 50\% in \citetalias{Cialone2018} paper. This discrepancy is again explained by the fact that in \citetalias{Cialone2018} all quantities were defined within $R_{\rm vir}$ and therefore contained the less virialized external regions. To be sure that this interpretation is correct, we analysed only for sake of this comparison also {\sc The Three Hundred} runs carried out with the same simulation code \textsc{gadget-music} as for the MUSIC clusters. We consider only $z=0$ clusters and evaluate all parameters within $R_{\rm vir}$. As such, it is recovered the same fraction of MUSIC relaxed clusters. The absence of a dynamical evolution is mainly due to the fraction in mass indicator ($f_{s,mm}$) used in \citetalias{Cialone2018} paper. This parameter significantly changes only when a great substructure enters into the dominant halo and does not consider all the other substructures as $f_s$ does (Fig.~\ref{fig:3D_dyn}). Therefore, for this work we prefer to consider $f_s$ because more sensitive to even minor mergers which are expected to perturb the ICM at the same level.

As shown in Fig.~\ref{fig:rel_com}, the criteria of \citet{300} return, instead, a much smaller percentage of `relaxed' objects for two main reasons: they consider as a factor also the energy virial ratio and they impose a stronger condition on $\Delta_r$ ($0.04$ versus $0.1$). As a result, less than 20\% of the clusters are now recognised as relaxed, with a redshift dependence similar to the one found here (Fig.~\ref{fig:3D_dyn}). 
Interestingly, using these criteria the results related to $R_{200}$ and $R_{500}$ are almost the same. We verified that the absence of an aperture dependency is due to the introduction of $\eta$: this parameter has an opposite dependency on the explored volume than the others. When we consider exclusively the $\eta$ parameter, we have more relaxed clusters at $R_{200}$ (63\% at $z=0$) than at $R_{500}$ (close to 34\%). This is due to the definition of $\eta$: it was initially introduced to study the dynamical state of isolated objects. Therefore, estimating it inside the clusters leads to other contribution in $E_s$ due to the interaction between the external regions of the clusters and the inner ones. Therefore this criterion could be used to restrict the analysis only to the "very relaxed" cluster sub-sample considering $R_{200}$, for which the hydrostatic assumption is more fulfilled.

\subsection{X-ray and y maps morphology} \label{ssec:res_Xy}

With the procedure described in Sections \ref{ssec:2D_morp} and \ref{ssec:2D_morp_eff}, we use the dynamical state classification results in the KS test to estimate the best aperture and the weights (Eq.~\eqref{eq:Wi}) of the morphological parameters for X-ray and $y$ maps. For the KS test, we calculate the parameters inside 4 equally spaced fractions of $R_{500}$, and in the case of the $c$ parameter, we consider 10 equally spaced inner radii varying in the range $[0.1-1] R_{\rm ap}$. The tuning procedure of X-ray and $y$ maps morphological indicators with the KS test returns small probability \textit{p}-values, but their variance changes of several orders of magnitudes over redshift. As an example, we have $p\sim10^{-7}$ ($z=0$) and $p\sim10^{-3}$ ($z=1$) for $y$ maps $P$ parameter, with an aperture of $R_{500}$. Although a decrease of performances is expected for most of the parameters for higher redshifts, we decided to use the median of KS-p value in \citetalias{Cialone2018} tuning procedure, instead of average, to be less affected by outlier and to obtain a single best aperture suitable for all the redshifts. In Table~\ref{tab:W_R} the best apertures from the KS-test tuning procedure and weights for the $M$ parameter are listed. The weights are expressed as fractions of $W_T=\sum_i W_i$, with $W_T=82$ for X-ray and $W_T=71$ for $y$ maps, to clearly show which parameter contributes more in $M$.

\begin{table}
\centering
\caption{Radii of the best apertures, in units of $R_{500}$, and relative weights ($W_i/\sum _iW_i$) of all the morphological parameters for X-ray and $y$ maps. In the case of the $c$ parameter, the two radii of the concentric apertures are listed.}
\label{tab:W_R}
\begin{tabular}{|c|c|c|c|c|c|c|}
\hline 
morphological & \multicolumn{3}{c|}{X maps} & \multicolumn{3}{c|}{$y$ maps} \\ 
\cline{2-7}
parameters & $R_{ap}$ & KS-p & $W$ & $R_{ap}$ & KS-p & $W$ \\ 
\hline 
$A$ & 0.50 & 8.6e-15 & 0.17 & 1.00 & 1.2e-11 & 0.15 \\ 
\hline 
\multirow{2}{*}{c} & 0.025 & \multirow{2}{*}{3.4e-20} & \multirow{2}{*}{0.24} & 0.05 & \multirow{2}{*}{6.6e-21} & \multirow{2}{*}{0.29} \\ 
& 0.25 & & & 0.25 \\
\hline 
$P$ & 0.25 & 5.3e-17 & 0.20 & 0.50 & 1.2e-9 & 0.13 \\ 
\hline 
$w$ & 0.75 & 1.2e-16 & 0.19 & 0.75 & 4.7e-19 & 0.26 \\ 
\hline 
$G$ & 0.50 & 4.7e-5 & 0.05 & 0.25 & 9.3e-4 & 0.04 \\ 
\hline 
$S$ & 1.00 & 4.1e-13 & 0.15 & 1.00 & 5.7e-10 & 0.13 \\ 
\hline 
\end{tabular}
\end{table}

Although most of the parameters are originally defined for X-ray observations, they show similar results also on $y$ maps. Two examples are the parameters $c$ and $w$ that weight even more in $y$ maps than in X-ray ones. The results of the KS analysis in the two sets of maps mostly differ for the identification of the best aperture to be used to compute the morphological indicators. The indicators that are influenced by small-scale variations favour a small aperture in the X-ray maps where the substructures are more evident. Indeed, some small central clumps might be in pressure equilibrium and therefore are hidden in the $y$ maps. Vice versa, the parameters that mostly consider large-scale inhomogeneity are better traced with a larger aperture in X-ray in order to capture more of the external signal, which is weaker with respect to the $y$ maps. Looking at the weights that each parameter carries, we notice that the $G$ parameter, as highlighted also by \citetalias{Cialone2018}, is the least effective parameter and it contributes to the combined parameter $M$ only three or four times less. This is because $G$ gives a global estimation of how much a cluster is prolate or oblate and does not take into account the detailed internal structure of ICM. Moreover, the spherical shape assumption is a rare case even for a relaxed cluster: most clusters are better described by an ellipsoidal shape and the $G$ parameter, thus, become less robust because it can strongly vary depending on the chosen projection.

\begin{figure}
    \includegraphics[width=\columnwidth]{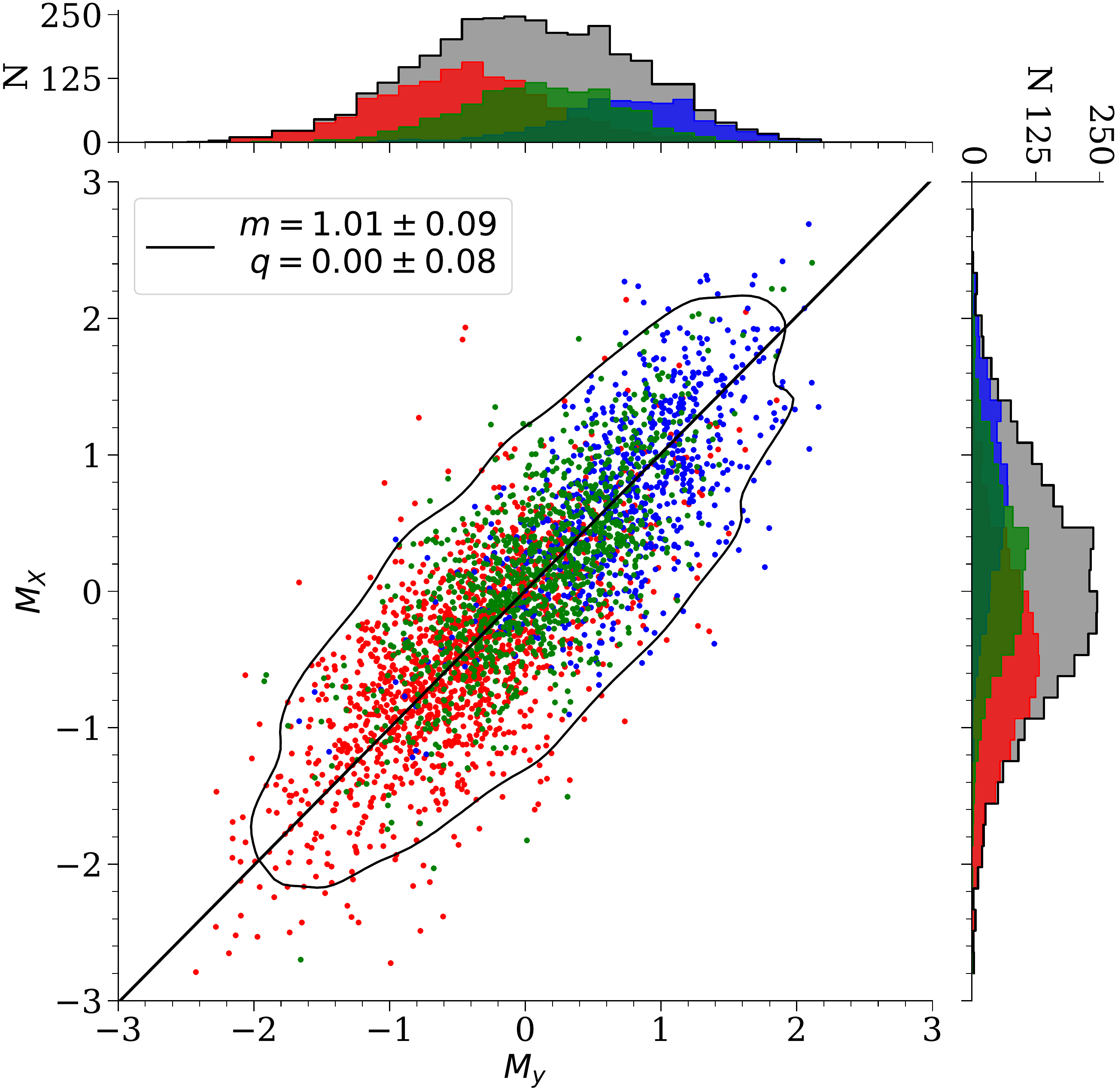}
    \caption{Scatter plot and marginal distributions of the relaxed (red), hybrid (green), and disturbed (blue) morphological combined $M$ parameter computed on X-ray and $y$ maps for all the 3240 clusters in the 10 redshift snapshots. A linear fit is also plotted in the figure, along with the $95\%$ confidence contour level of the data distribution. The result of the fit is shown in the legend of the figure.}
    \label{fig:M}
\end{figure}

\begin{figure}
    \includegraphics[width=\columnwidth]{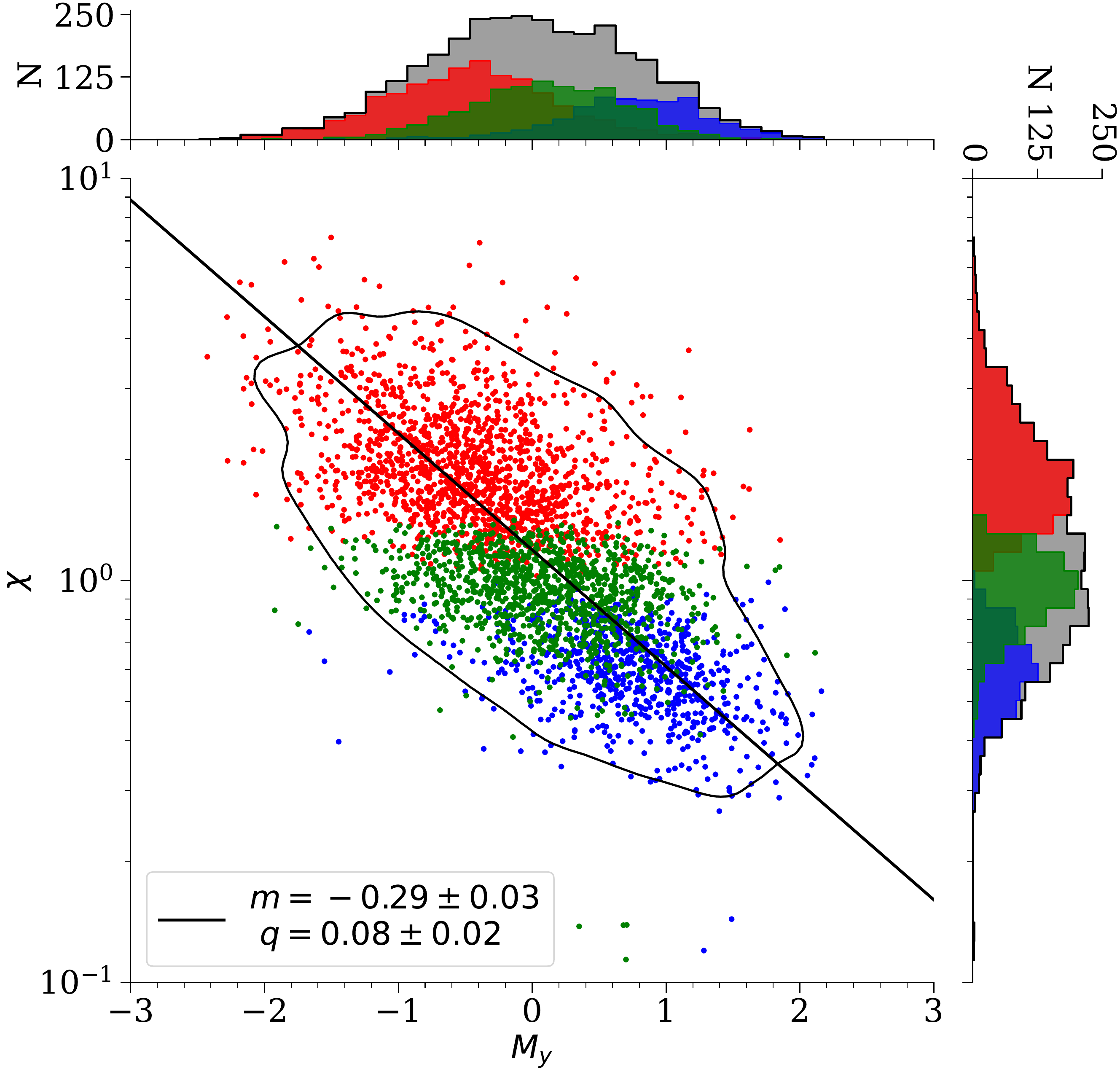}
    \caption{Scatter plot, in semilog scale, between the dynamical $\chi$ indicator and the morphological $M$ parameter for $y$ maps, with their distributions in the marginal plots. The straight line in this figure corresponds to a linear fit of the data and red, green, blue colours are associated with the relaxed, hybrid and disturbed dynamical classes. The $95\%$ confidence level of the data is also shown in the figure. The slope and the vertical intercept of the linear fit are listed in the legend of the figure. Similar results are obtained also for the X-ray $M$ parameter.}
    \label{fig:XvsM}
\end{figure}

A linear correlation between the combined parameters estimated in the two maps ($M_y$ and $M_X$) is present, as shown in Fig.~\ref{fig:M}, with a Spearman correlation coefficient of $\rho=0.80$. From this figure and hereafter, all the fits are performed with a least-square algorithm to reduce the effect of outliers or leverage points present in the data. For the $M_y$-$M_X$ relation, the best fit is in agreement with the line of equality, indicating that, on average, this set of indicators works efficiently both on X-ray or $y$ maps. From the marginal distributions in the top and side panels of Fig.~\ref{fig:M}, the most common clusters in our sample are morphological hybrid clusters, since the peaks of the $M$ distributions are close to $M\sim0$, while the two tails are associated with the more relaxed (negative tail) and more disturbed (positive) clusters. It is relevant to stress how both these distributions are remarkably similar to the $\chi$ distribution already shown in Fig.~\ref{fig:3D_dyn}. The relation between the dynamical and morphological state is shown in Fig.~\ref{fig:XvsM}, where the scatter plot between the two parameters, $\chi$ and $M$, is studied. The Spearman coefficient indicates a relatively strong correlation: $\rho\sim-0.66$, using $M$ from either $y$ or X-ray maps. As a result, the M parameter derived from these maps could be used as a single good proxy of the dynamical state of galaxy clusters, since it represents a dynamical state weighted combination of different morphological aspects of clusters. In Fig.~\ref{fig:M} and Fig.~\ref{fig:XvsM} we use colours to distinguish the three classes defined in Sect.~\ref{sec:DS}: relaxed, hybrid, and disturbed. We intend to show how a rigid morphological classification based on the thresholds of these morphological parameters, in order to infer the dynamical state, could lead to contamination by other classes. Considering the dynamical classification in $M$ distribution, the hybrid clusters in the figures show, in fact, a non-negligible superimposition over the other classes. Instead, the relaxed and disturbed distributions are enough separated, as also highlighted by the low KS-p values in Table~\ref{tab:W_R}, returned by the tuning procedure. To quantify what is visually represented in the figures, the median values, with 16th and 84th percentiles of the three dynamical state distributions of $M$, are summarised in Table~\ref{tab:M}. The superimposition between the three distributions also depends mildly on $z$, since in the past clusters are less relaxed than at $z=0$. Looking at low and high redshift clusters, the median in Table~\ref{tab:M} of the relaxed, hybrid and disturbed distributions are closer among each other at $z=1$ than at $z=0$, where hybrid and disturbed medians move towards higher values of $M$. However, the large spread of $M$ values described by the percentiles in Table~\ref{tab:M} do not show any statistical significant redshift evolution of $M$ parameter.

\begin{table}
\centering
\caption{Median of relaxed, hybrid and disturbed distributions for the $M$ parameter, at $z=0$, $z=1$, and considering all the redshift. The first rows correspond to the $y$ maps, while the second ones to the X-ray maps. The 16th and 84th percentiles are also listed in the table, with a $\pm$ near the median values.}
\label{tab:M}
\begin{tabular}{|c|c|c|c|}
\hline 
$M$ & $z=1$ & $z=0$ & all \\
\hline 
\multirow{2}{*}{Relaxed} & $-0.50^{+0.53}_{-0.69}$ & $-0.50^{+0.76}_{-0.65}$ & $-0.45^{+0.62}_{-0.66}$ \\
& $-0.42^{+0.78}_{-0.83}$ & $-0.48^{+0.66}_{-0.58}$ & $-0.48^{+0.69}_{-0.68}$ \\ 
\hline 
\multirow{2}{*}{Hybrid} & $-0.06^{+0.59}_{-0.44}$ & $0.34^{+0.45}_{-0.66}$ & $0.10^{+0.58}_{-0.59}$ \\
& $-0.08^{+0.60}_{-0.54}$ & $0.28^{+0.74}_{-0.62}$ & $0.14^{+0.65}_{-0.60}$ \\ 
\hline 
\multirow{2}{*}{Disturbed} & $0.63^{+0.50}_{-0.61}$ & $0.84^{+0.57}_{-0.58}$ & $0.78^{+0.50}_{-0.55}$ \\
& $0.67^{+0.57}_{-0.91}$ & $0.73^{+0.78}_{-0.67}$ & $0.74^{+0.64}_{-0.68}$ \\
\hline 
\end{tabular}
\end{table}

\begin{figure*}
	\includegraphics[width=\columnwidth]{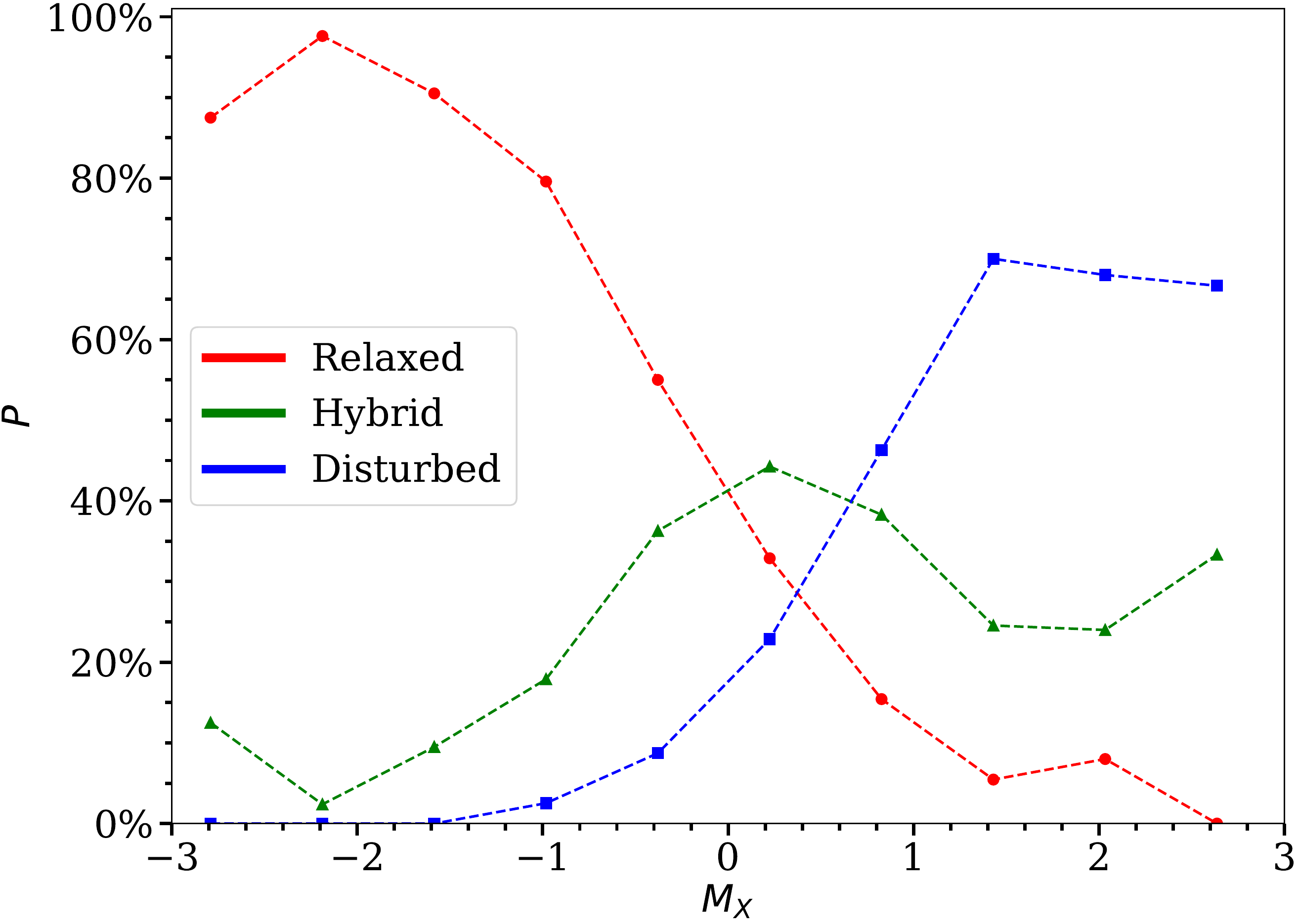}
	\includegraphics[width=\columnwidth]{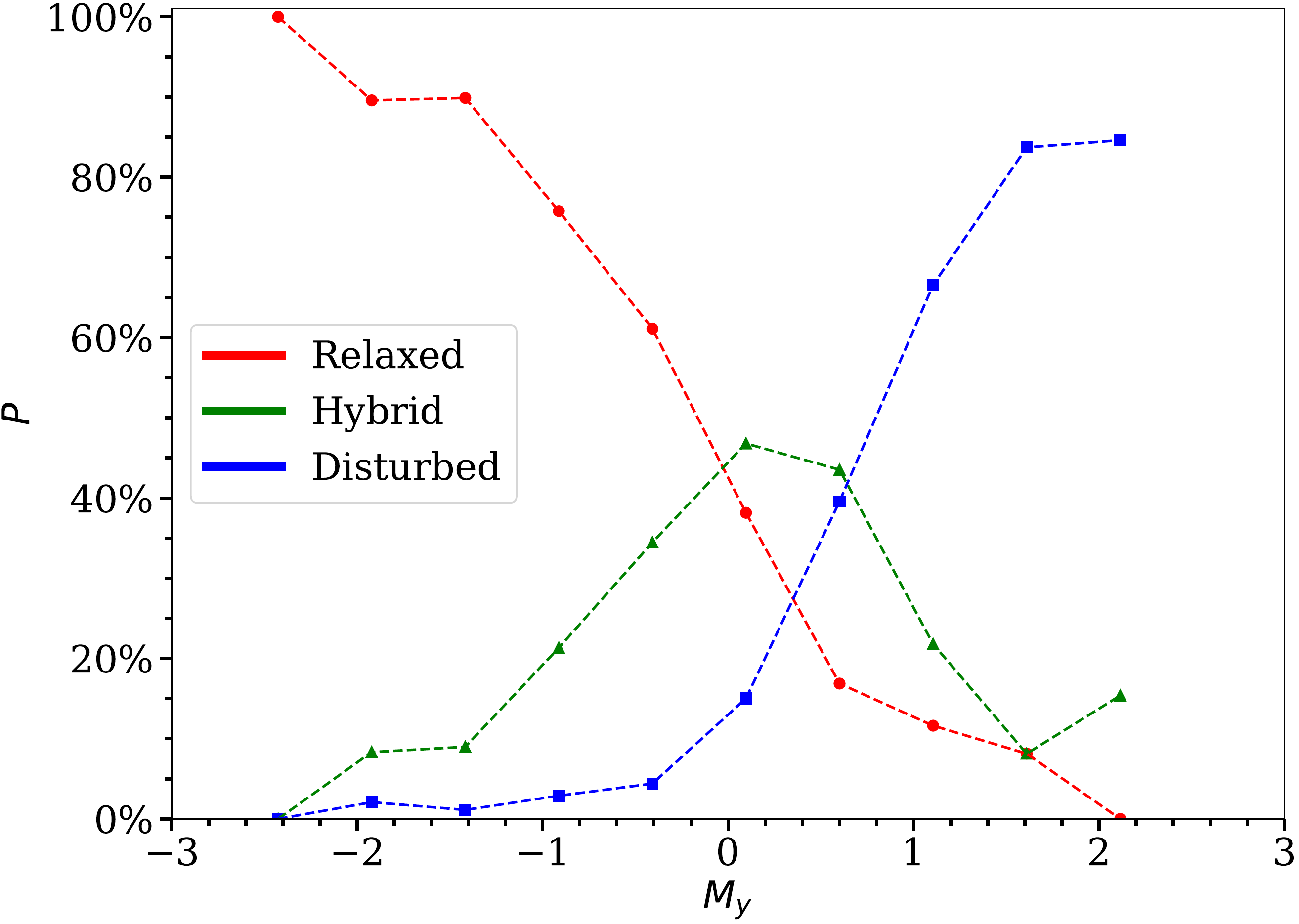}
    \caption{Probability functions $P$ applied to the combined morphological parameter $M$ for X-ray (left panel) and $y$ (right panel) maps, as defined in Eq.\eqref{eq:IG_rd} (Sec.~\ref{ssec:2D_morp_eff}), for all the 3240 studied clusters in the sample. The three colours correspond to the dynamical classes of relaxed (red), hybrid (green) and disturbed (blue) clusters.}
    \label{fig:IG}
\end{figure*}

An even more effective way to quantify the level of contamination is to perform a ROC analysis, introduced in Sec.~\ref{ssec:2D_morp_eff} and largely employed in Appendix \ref{sec:A_ROC}. In particular, in the upper left panel of Fig.~\ref{fig:ROC} we use this analysis to highlight how the morphological parameter $M$ can effectively separate the dynamical classes defined through $\chi$ and, specifically, the relaxed clusters from the disturbed (blue), the hybrid (green), and from all the non-relaxed (grey), e.g. disturbed and hybrid. Considering only the disturbed cluster, the associated ROC curve is closer to the perfect case (with $AUC\sim0.9$ meaning that the two classes are very well separated) than if we consider only the hybrid ($AUC\sim0.74$) or taking hybrid and disturbed clusters together ($AUC\sim0.80$). Therefore the contamination is mainly due to hybrid clusters when a threshold is chosen. From the maximum values of $MCC$ and $J$, the two statistics give similar thresholds for $M$, all close to the expected separation value of $0$: $M_{MCC}=-0.02$, $M_{J}=-0.01$ for $y$ maps and $M_{MCC}=-0.13$, $M_{J}=-0.04$ for X-ray ones. In Tables~\ref{tab:Purity_y} and \ref{tab:Purity_X}, the purity $p$ and the completeness $C$ of the relaxed and non-relaxed sub-samples are shown, for a set of $M$ values. In Fig.~\ref{fig:IG} the merit function $P$ defined in Eq.\eqref{eq:IG_rd} (Sec.~\ref{ssec:2D_morp_eff}) is shown with respect to the $M$ parameters considering all the simulated clusters. To calculate $P$, we divide $M$ values into ten equally spaced bins between their minimum and maximum values. The corresponding values of $P$ for X-ray and $y$ maps are shown separately in the two panels of Fig.~\ref{fig:IG}. Considering that negative values of $M$ are associated with relaxed clusters and disturbed to positive ones, it is not surprising that the trends of the relaxed and disturbed clusters are opposite and reach their maximum in the extreme values. Without the hybrid clusters, the contamination of the disturbed clusters is contained: at $M\sim0$ it is at the level of $\sim13$-$20\%$ and lower for $M<0$. Therefore, hybrid clusters represent the major source of contamination for relaxation definition, with $P$ close to $45\%$ at $M\sim0$ and $P\sim20\%$ at $M\sim-1$.

\begin{table}
\centering
\caption{Purity and completeness, in percentage, for a selected sample of $M$ values estimated in $y$ maps for the relaxed and non-relaxed (disturbed plus hybrid) classes.}
\label{tab:Purity_y}
\begin{tabular}{|c|c|c|c|c|}
\hline
 & \multicolumn{2}{c|}{Relaxed} & \multicolumn{2}{c|}{Non-relaxed} \\ 
\hline 
$M_y$ & $p$ & $C$ & $p$ & $C$ \\ 
\hline 
-2   & 100   & 1.04  & 22.45 & 100  \\ 
\hline 
-1.5 & 90.48 & 5.29  & 22.88 & 99.56  \\ 
\hline 
-1   & 87.38 & 19.76 & 24.70 & 97.73  \\ 
\hline 
-0.5 & 78.33 & 46.90 & 29.62 & 89.63  \\ 
\hline 
 0   & 68.00 & 77.04 & 41.14 & 70.99  \\ 
\hline 
0.5  & 56.64 & 92.28 & 56.73 & 43.59  \\ 
\hline 
 1   & 48.36 & 97.70 & 75.07 & 16.86  \\ 
\hline 
1.5  & 45.12 & 99.72 & 84.38 & 3.33  \\ 
\hline 
 2   & 44.43 & 100   & 83.33 & 0.33  \\ 
\hline 
\end{tabular} 
\end{table}

\begin{table}
\centering
\caption{Purity and completeness, in percentage, for a selected sample of $M$ values estimated in X-ray maps for the relaxed and non-relaxed (disturbed plus hybrid) classes.}
\label{tab:Purity_X}
\begin{tabular}{|c|c|c|c|c|}
\hline
 & \multicolumn{2}{c|}{Relaxed} & \multicolumn{2}{c|}{Non-relaxed} \\ 
\hline 
$M_X$ & $p$ & $C$ & $p$ & $C$ \\
\hline 
-2   & 93.75 & 2.02 & 22.57 & 99.89 \\ 
\hline 
-1.5 & 92.98 & 7.38 & 23.16 & 99.56 \\ 
\hline 
-1   & 88.73 & 21.92 & 24.96 & 97.78 \\ 
\hline 
-0.5 & 79.52 & 48.36 & 29.63 & 90.02 \\ 
\hline 
 0   & 66.36 & 75.37 & 38.68 & 69.55 \\ 
\hline 
0.5  & 55.87 & 92.35 & 53.06 & 41.71 \\ 
\hline 
 1   & 49.18 & 97.91 & 65.17 & 19.30 \\ 
\hline 
1.5  & 45.76 & 99.51 & 73.91 & 5.99 \\ 
\hline 
 2   & 44.58 & 99.86 & 66.67 & 1.05 \\ 
\hline 
\end{tabular} 
\end{table}

\subsection{Morphological offset parameters results} \label{ssec:2D_off_res}

The BCGs, estimated from r optical band maps of the clusters described in Sec.~\ref{ssec:maps}, have been identified as the most luminous galaxy inside an aperture of $0.5R_{500}$ centred on the maximum density peak of the clusters. This is done in order to reduce the selection error of a BCG gravitationally bound to a substructure present in the outskirt of the clusters. We recall that the pixel resolution is fixed in arcsec, one pixel in physical units span from $0.75$ to $6.67$ kpc moving from $z=0$ and $z=1$. These limits are, on average, smaller than $0.05$ and $0.5$ per cent of $R_{500}$.

\begin{figure}
    \includegraphics[width=\columnwidth]{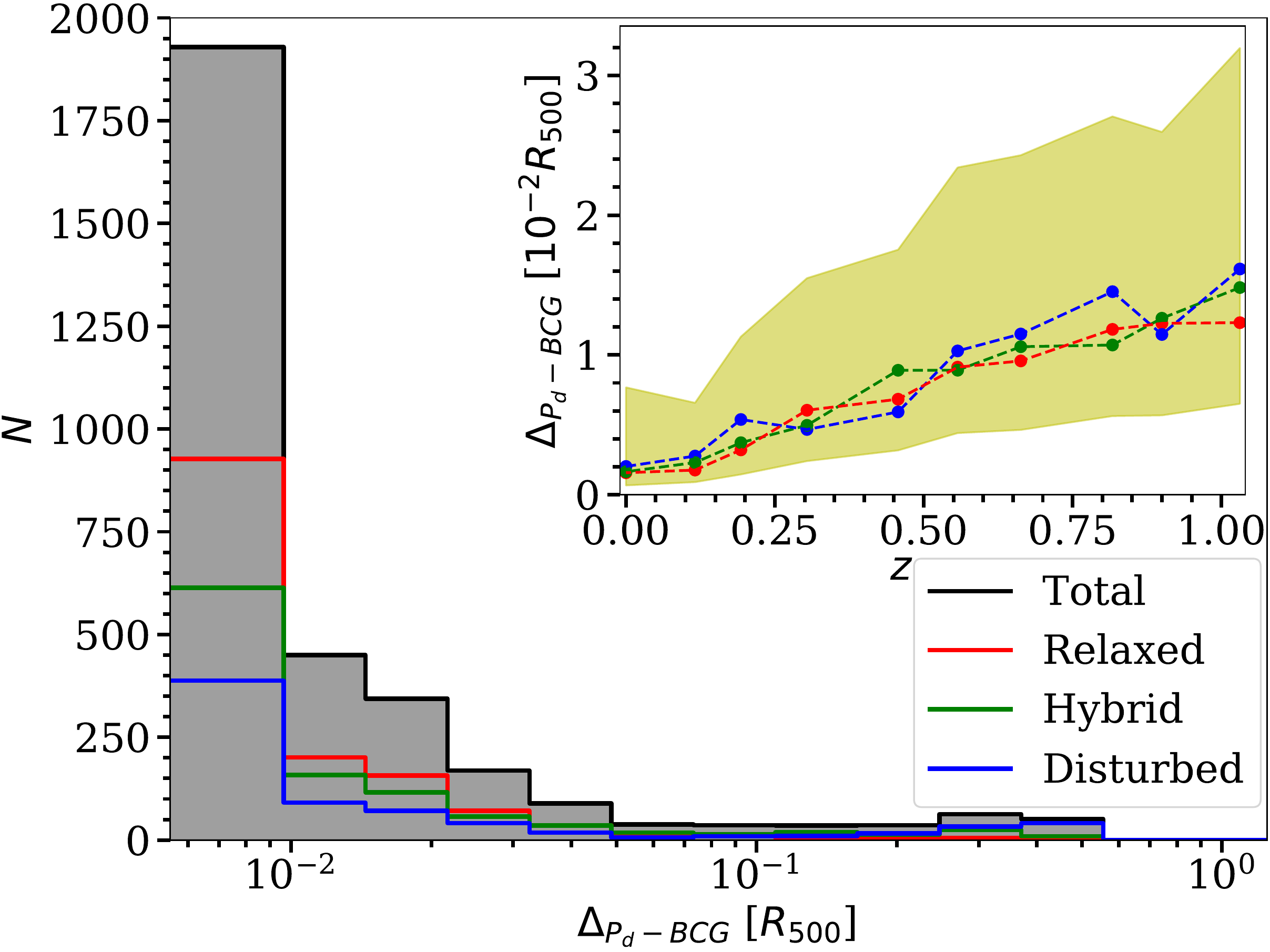}
    \caption{Histogram, in log scale, of the BCG-density peak offset, $\Delta_{P_d-BCG}$, and the median values of the distributions across the redshift in the inset figure. $\Delta_{P_d-BCG}$ is in units of $R_{500}$ and red, blue, green, and black lines correspond respectively to the relaxed, disturbed, hybrid and total distributions. The left limit in the histograms corresponds to the maximum resolution limit of the optical maps. The yellow area in the inset plot corresponds to the values between the 16th and 84th percentile of the distributions.}
    \label{fig:BCG_dP}
\end{figure}

In Fig.~\ref{fig:BCG_dP} the distribution of the offset between the BCG position and the theoretical centre, $\Delta_{D_P-BCG}$, is shown. Independently of the redshift, almost 92\% of the clusters in the simulation presents an offset $\Delta_{D_P-BCG} < 0.05R_{500}$, while the large majority has an offset below $0.02R_{500}$ (see the insert of the figure). The clusters with a large offset are associated with objects with low values of $\chi$ or large values of $M$ and typically are classified as non-relaxed objects. Only 3\% of relaxed clusters present large offsets due to relaxation processes still in action, highlighted by irregular morphology ($M>0$) or with a farther, slightly brighter galaxy than the one nearest to the peak density. Another result concerning the BCG position is its variation with the redshift. Considering the median of the offset distribution, it linearly decreases by a factor of 5 from high redshift to lower redshift clusters, independently on the dynamical state as also shown in the inset in Fig.~\ref{fig:BCG_dP}. The yellow area in the inset corresponds to the 16th and 84th percentile of the distributions.

We can conclude that, in general, the BCG position in {\sc The Three Hundred} cluster sample does not depend strongly on the dynamical state and its position can be used in observations as a good tracer of the total density peak of galaxy clusters, except for a few ($\sim 8\%$) disturbed clusters. A detailed discussion of this topic is beyond the goal of this paper, but the BCG Paradigm is weakly fulfilled, since most BCGs in {\sc The Three Hundred} sample are close to the density peak, but not completely at rest. For a more detailed study of this topic, see the papers of \citet{Coziol2009, Cui2016, Harvey2017, Lopes2018, DePropris2020}.

\begin{figure*}
    \includegraphics[width=\columnwidth]{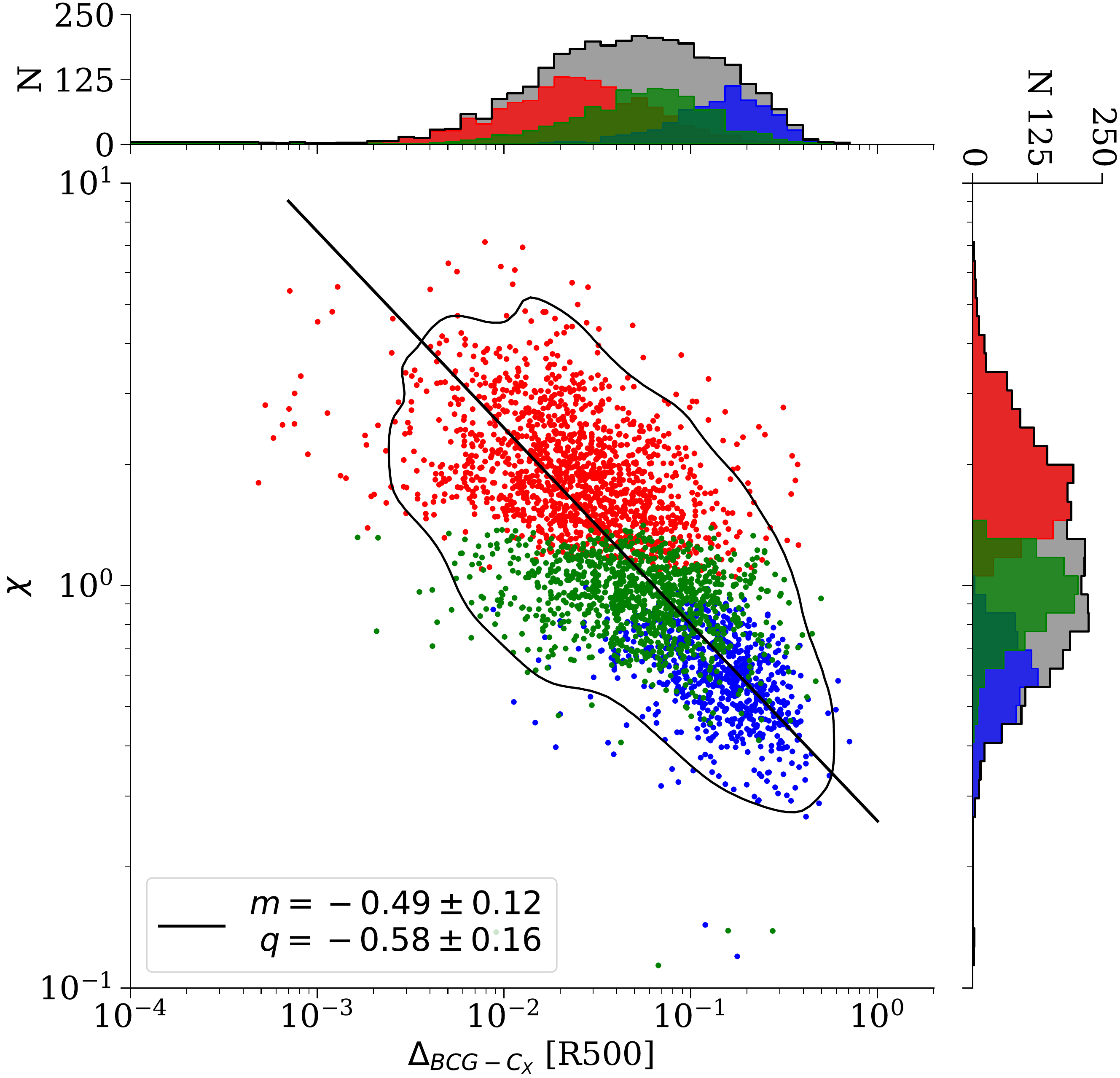}
    \includegraphics[width=\columnwidth]{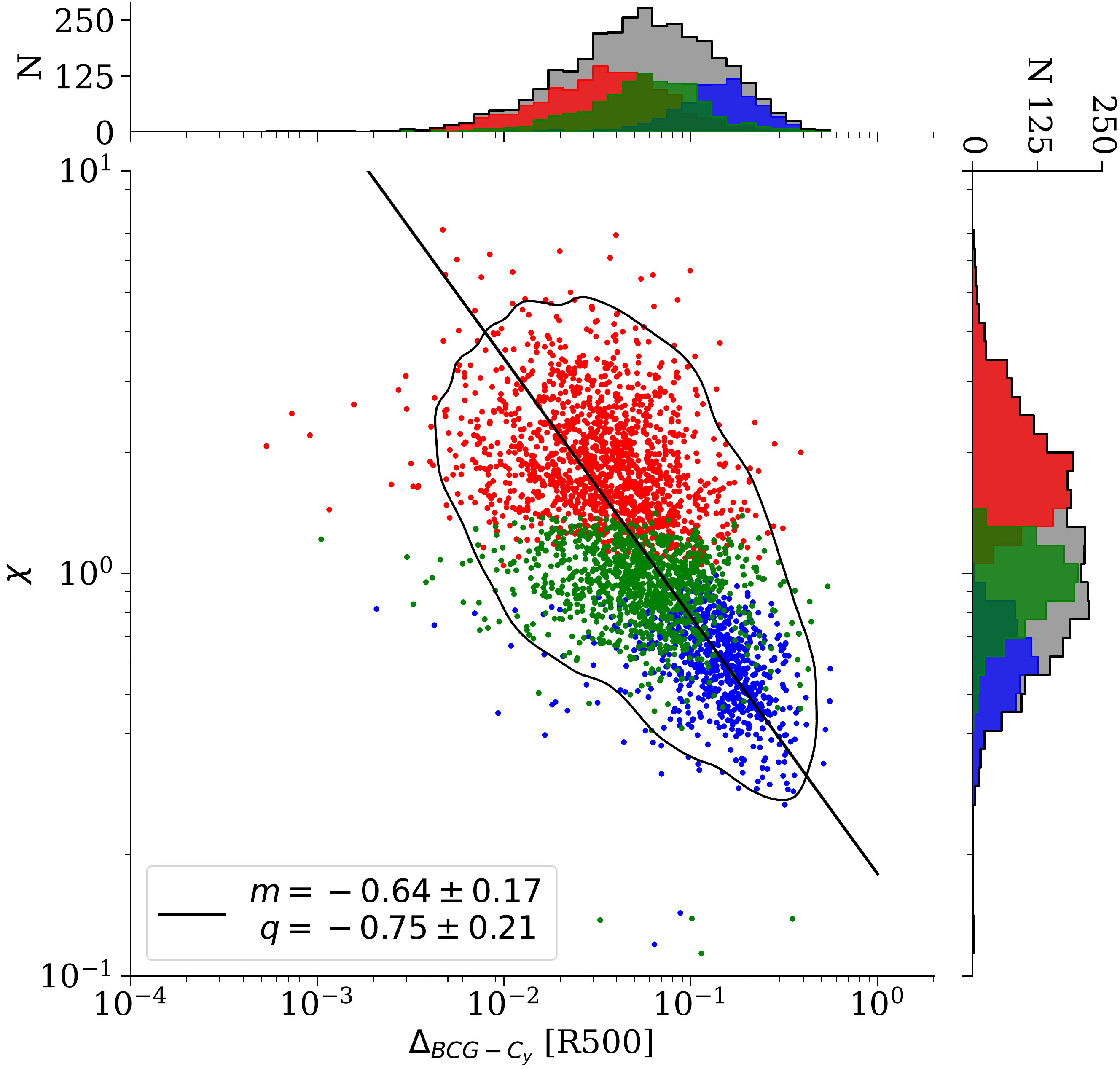}
    \includegraphics[width=\columnwidth]{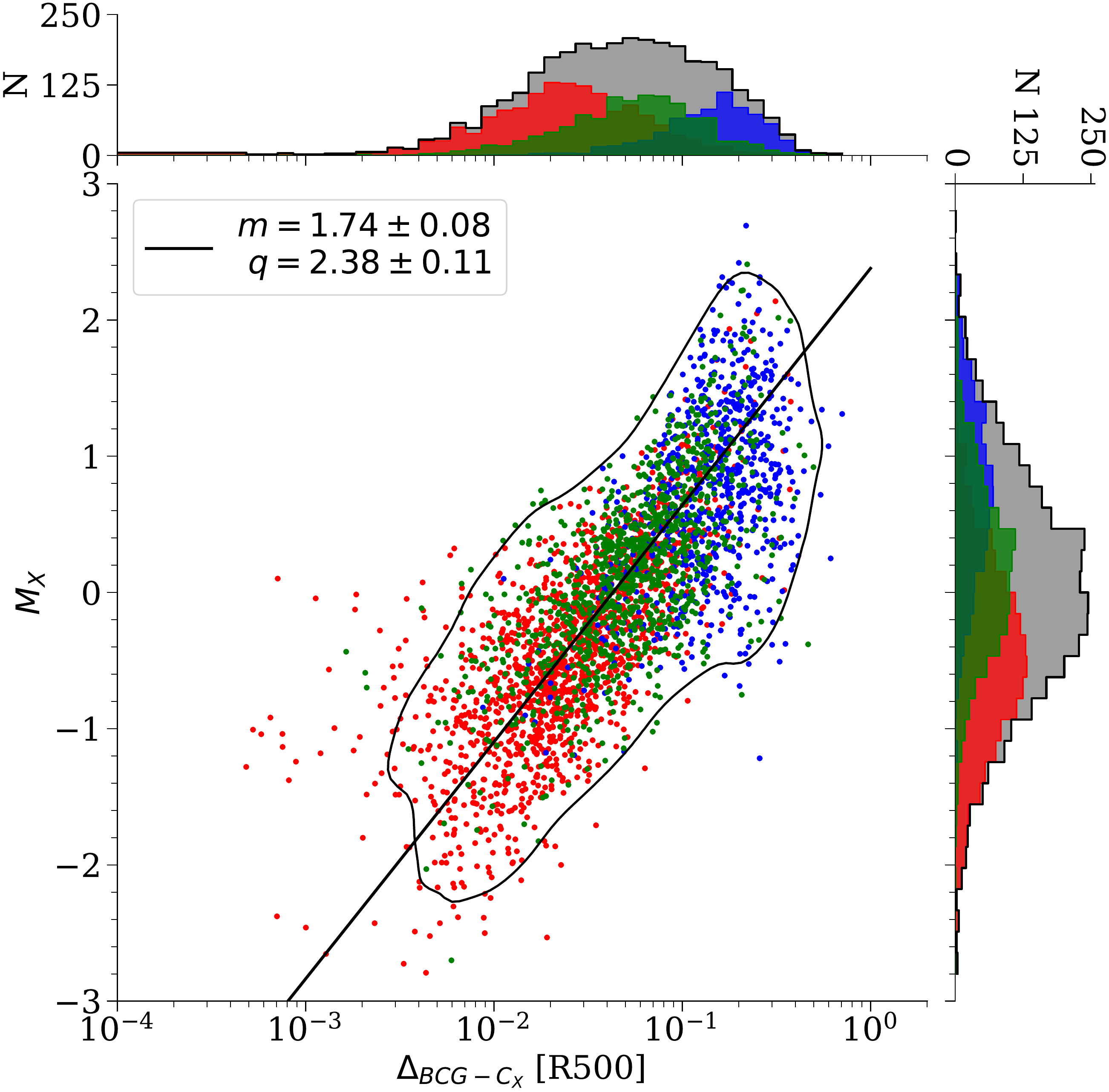}
    \includegraphics[width=\columnwidth]{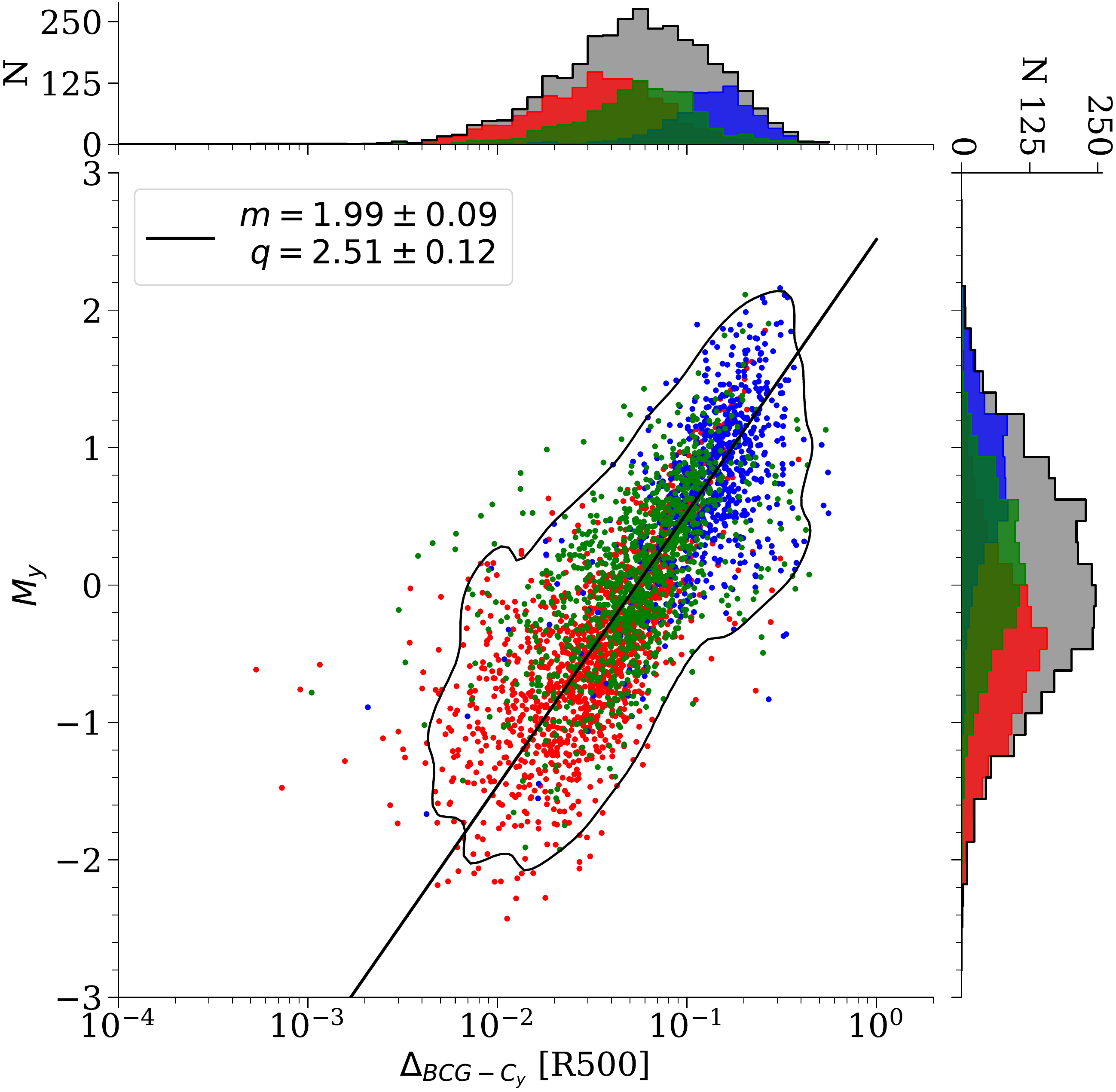}
    \caption{Scatter plots and distributions for the offset parameter, $\Delta$, between the BCG and the X-ray (left column) or $y$ (right) centroids. The offset parameters are calculated in terms of $R_{500}$ and considering all the 3240 clusters at the different redshifts. The upper row shows the correlation, in log-log scale, between the two offset parameters and the dynamical $\chi$ indicator defined in Sec.~\ref{sec:DS}. The lower row shows the result of the comparison between the offset parameters and the combined morphological $M$ parameter for X-ray (left) and $y$ maps. Red, green, blue colours are associated with relaxed, hybrid and disturbed clusters. The best fit and the 95\% confidence level of the data distribution are also shown in the figures, with the slope and y-intercept listed in the legends of the panels.}
    \label{fig:correlation_off}
\end{figure*}

We move now to compute the offsets between the BCG centre and the ICM centres identified as ICM peaks or as ICM centroids from the X-ray and $y$ maps (see Sec.~\ref{ssec:maps}). As expected, the offsets with respect to the positions of the peaks, $\Delta_{BCG-P_{X,y}}$, show no clear correlation with the dynamical state. Furthermore, the ROC curves present in the lower right panel of Fig.~\ref{fig:ROC} in the Appendix confirm their inefficiency: all the curves are close to the random guess line, with a median average AUC close to $0.61$, for the binary test of the relaxed population against the non-relaxed one. In absence of strong inhomogeneities or disturbances in the ICM \citep[like major merger events, as for cluster A370,][]{Molnar2020}, the ICM peaks are almost coincident with the total density peaks used before. Therefore both the map peaks and the BCG position are good estimates of the cluster centre.

However, regarding the positions of the centroids, relaxed, disturbed and hybrid clusters show different offsets with respect to all centres discussed before (the theoretical one or the total density peak, the ICM peaks and the BCG position). The centroids results are more affected by inhomogeneities, or more in general by disturbances, in the overall ICM structure.
The offset between the BCG and the two centroids, $\Delta_{BCG-C_{X,y}}$, shows a relatively strong correlation with the dynamical state $\chi$ indicator: $\rho=-0.63$ for $y$ centroid and $\rho=-0.69$ for X-ray maps. Moreover, these offsets have also a high correlation with $M$: $\rho=0.80$ for $y$ maps and $\rho=0.79$ for X-ray maps. In Fig.~\ref{fig:correlation_off} the scatter plots and the distributions of these indicators respective to $\chi$ and $M$ are shown. These results are corroborated by the performance analysis of ROC curves. The centroid parameters detach from the random guess, showing for $\Delta_{BCG-C_{X}}$ an $AUC\sim0.83$ ($0.78$ for $\Delta_{BCG-C_{y}}$) if we consider the non-relaxed class and $AUC\sim0.94$ ($0.93$) considering instead only the disturbed clusters. As expected, there are slightly better results with X-ray data, which again emphasise the presence of even small substructures, with AUC values that are generally larger than $y$ counterparts. In Fig.~\ref{fig:ROC} in the Appendix, the ROC curves for the offset parameters $\Delta_{BCG-C_{X}}$, $\Delta_{BCG-P_{X}}$, and $\Delta_{P_X-C_{X}}$ are shown. The other possible offsets of the centroids with the other tracers, as the peak positions, show similar results to the BCG one. For the thresholds on BCG-centroids offset parameters, the maximum of $J$ and $MCC$ are slightly different for X-ray and $y$ data. For the offsets with X-ray centroid, the suggested threshold is at $0.05\times R_{500}$ while for $y$ centroid it is at $0.06$-$0.07\times R_{500}$.

\subsection{Comparison with the observational estimates of relaxed fraction of galaxy clusters} \label{ssec:2D_obs_comp}

The estimation of the fraction of relaxed galaxy clusters has been extensively studied in observations. However, a direct comparison is not straightforward for the results obtained with different morphological parameters and based on different samples, as highlighted by \citet{Cao2021}. Morphological parameters are often differently defined depending on the main topic of the paper or the limitation of the analysis procedure. Furthermore, the comparison between different clusters samples could be affected by selection effects, as the Malmquist bias, especially for flux-limited X-ray samples \citep{Hudson2010, Chon2017}. In fact, \citet{Rossetti2017, AndradeSantos2017, Chon2017} have highlighted the presence of a bias between the SZ and X-ray sample of cool core (CC) and non-cool core (NCC) fractions and in the relaxed cluster fraction \citep[$52\pm4\%)$ vs $\sim74\%$][]{Rossetti2016}. In SZ-selected clusters, \citet{Lopes2018} have found a higher fraction of substructures than X-ray selected clusters. \citet{Jeltema2008} and \citet{Maughan2008} found with numerical simulations and observations a redshift evolution of the dynamical state. Instead, \citet{Bartalucci2019} found a weak evolution of their combined morphological parameter with $z$, while \citet{Nurgaliev2017} and \citet{McDonald2017} found no significant statistical difference using photon asymmetry, $A_{phot}$, and centroid shift parameters in describing X-ray morphology of X-ray and SZ selected clusters samples used, over the explored redshift range. Therefore, the number of clusters classified as relaxed varies significantly in the literature according to the different samples or morphological parameters used in each paper.

\begin{table*}
\centering
\caption{The fraction of relaxed clusters for different observational samples available in the literature, compared with {\sc The Three Hundred} results. The total number of objects and the redshift ranges of the cluster samples are also reported in the table. The (*) symbol denotes that these works do not explicitly quote the fraction of relaxed clusters, so we derived them from their tables or figures. For {\sc The Three Hundred} sample, we excluded the fraction of relaxed clusters from the $y$ centroid-BCG positions offset parameter, using the $MCC$ threshold. It is equal to $57\%$ due to the flatness of the $MCC$ peak.}
\label{tab:relax_fraction}
\begin{tabular}{|c|c|c|c|c|}
\hline
\multirow{2}{*}{Paper} & \multirow{2}{*}{Number of objects} & relaxed cluster & \multicolumn{2}{c|}{redshift range} \\
 & & fractions [\%] & $z_{min}$ & $z_{max}$ \\
\hline
{\sc The Three Hundred} & $3240$ & $44$-$49$ & - & $<1.031$ \\ 
\hline
\citet{Santos2008} (low z) & $11$ & $64$ & $0.15$ & $0.3$\\
\citet{Santos2008} (high z) & $15$ & $73$ & $0.7$ & $1.39$\\
\hline
\citet{Sanderson2009} & $65$ & $37$ & $0.15$ & $0.3$  \\ 
\hline
\citet{Zhang2010}* & \multirow{2}{*}{$12$} & \multirow{2}{*}{$42$} & \multirow{2}{*}{$0.15$} & \multirow{2}{*}{$0.3$}  \\
or \citet{Okabe2010}* & &  &  & \\
\hline
\citet{Cassano2010}* & $32$ & $44$ & $0.2$ & $0.4$  \\ 
\hline
\citet{Bohringer2010}* & $31$ & $48$ & $0.06$ & $0.18$  \\ 
\hline
\citet{Mann2012}* & $108$ & $44$ & $0.15$ & $0.7$  \\ 
\hline
\citet{Maughan2012}* & $114$ & $18-25$ & $0.1$ & $1.3$  \\ 
\hline
\citet{Mahdavi2013}* & $50$ & $26-28$ & $0.15$ & $0.55$  \\ 
\hline
\citet{Nurgaliev2013} & $36$ & $33$ & $0.3$ & $0.9$  \\ 
\hline
\citet{Parekh2015} & $84$ & $23$ & $0.02$ & $0.9$  \\ 
\hline
\citet{Mantz2015} & $361$ & $16$ & $0.05$ & $1.2$  \\ 
\hline
\citet{Lavoie2016} & $85$ & $65$ & $0.043$ & $1.05$\\
\hline
\citet{Rossetti2016} (SZ sample) & $132$ & $52\pm4$ & $0.02$ & $0.87$\\
\hline
\citet{Rossetti2017} (SZ sample) & $169$ & $29\pm4$ & $0.04$ & $0.87$\\
\citet{Rossetti2017} (X-ray) & $104$ & $59\pm5$ & $0.15$ & $0.7$\\
\hline
\citet{AndradeSantos2017} (SZ sample) &  $164$ & $28$-$39$ & - & $<0.35$ \\
\citet{AndradeSantos2017} (X-ray) &  $100$ & $44$-$64$ & $0.025$ & $0.3$ \\
\hline
\citet{Lovisari2017}* &  $120$ & $32$ & $0.01$ & $0.55$ \\
\hline
\citet{Chon2017} (Volume limited sample) &  $93$ & $29$ & - & $<0.1$ \\
\citet{Chon2017} (Flux limited sample 1) &  $51$ & $41$ & - & $<0.1$ \\
\citet{Chon2017} (Flux limited sample 2) &  $42$ & $43$ & - & $<0.1$ \\
\hline
\citet{Lopes2018} (X-ray sample, optical indicators) &  $62$ & $47$-$66$ & $0.01$ & $0.1$ \\
\citet{Lopes2018} (X-ray sample, X-ray indicators) &  $62$ & $65$-$69$ & $0.01$ & $0.1$ \\
\citet{Lopes2018} (SZ sample, optical indicators) &  $40$ & $38$-$63$ & $0.01$ & $0.1$ \\
\citet{Lopes2018} (SZ sample, X-ray indicators) &  $40$ & $48$-$53$ & $0.01$ & $0.1$ \\
\hline
\citet{Bartalucci2019}* & $74$ & $46$ & $0.08$ & $1.13$ \\ 
\hline
\citet{Zenteno2020} & $288$ & $14$ & $0.1$ & $0.9$ \\ 
\hline
\citet{Yuan2020} & $964$ & $51.2$ & $0.003$ & $1.75$ \\ 
\hline
\end{tabular}
\end{table*}

We summarise several values for the fractions of relaxed galaxy clusters that can be found in the literature in Table~\ref{tab:relax_fraction} and compare them with the results shown in this work. Considering the cool-core clusters as relaxed clusters, we also list their fractions obtained from thresholds on the concentration ratio parameter $c$. For some of the works listed, the authors do not specify the number of relaxed clusters from the used morphological indicators. The fractions in Table~\ref{tab:relax_fraction} are then calculated according to their criteria, considering the data present in their tables or figures. These values are marked in the table with a (*), near the reference. We further note that this unbiased selection merely includes all the works with different observations, methods, criteria and thresholds. No normalisation nor correction is included.

\begin{figure}
    \includegraphics[width=\columnwidth]{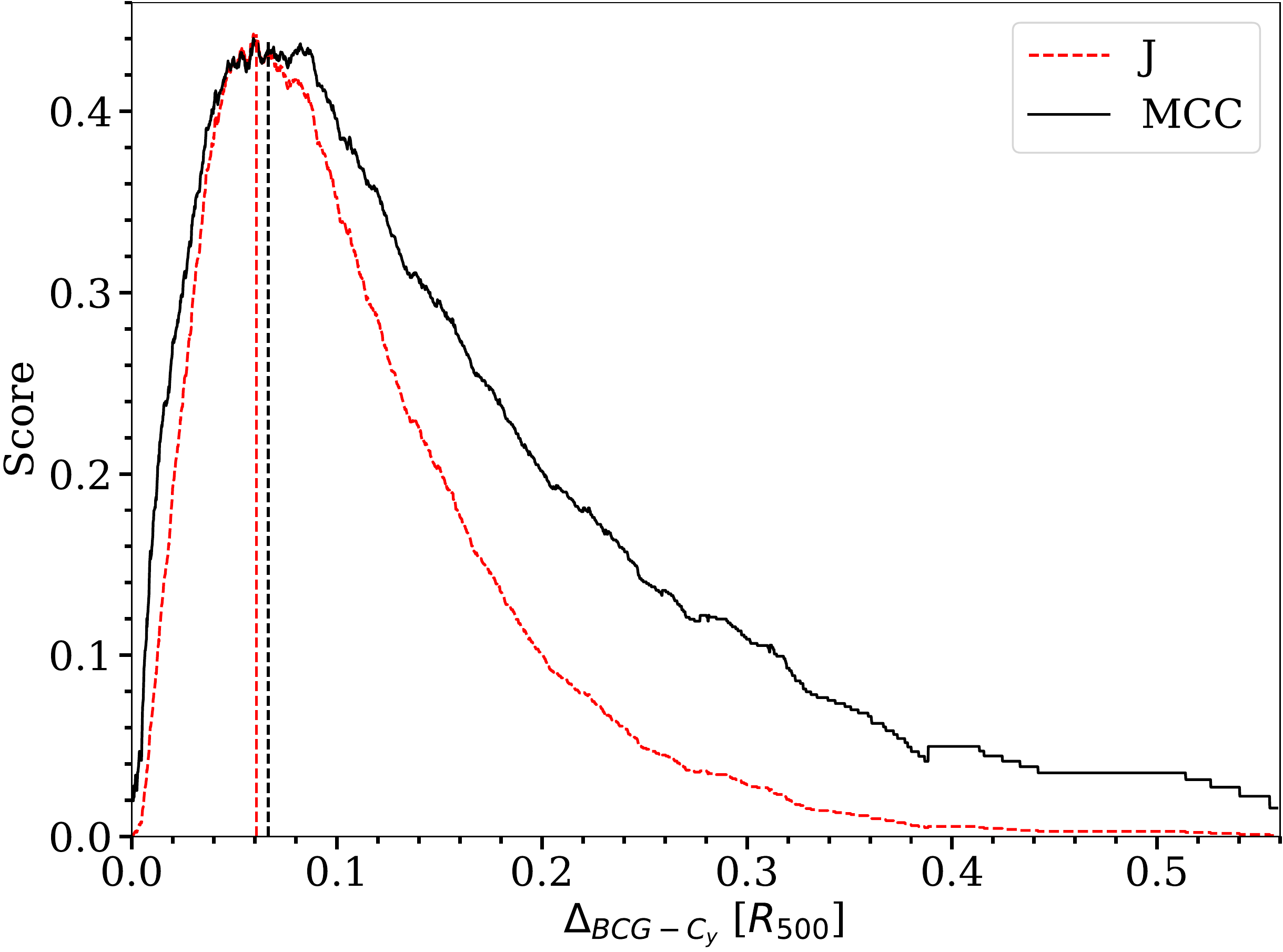}
    \caption{Distribution of $MCC$ (black solid line) and $J$ (red dashed line) merit functions, defined in Sec.~\ref{ssec:2D_morp_eff}, for the offset parameter between BCG and $y$ centroids, in units of $R_{500}$. The inferred thresholds on the morphological parameter are represented as vertical lines in the figure.}
    \label{fig:MCC_J}
\end{figure}

The relaxed cluster fraction that we can infer from morphological parameters in our simulated sample is close to \citet{Bartalucci2019} ($\sim46\%$) or SZ clusters of \citet{Rossetti2016} ($\sim52\%$) and \citet{Lopes2018} ($49\pm8$) for the morphological indicator based on the offsets. If we apply the thresholds from $MCC$ and $J$ to the combined parameter $M$, we recover a fraction of $44$-$49\%$ of relaxed clusters, while from the dynamical analysis in Sec.~\ref{ssec:3D_ds}, the total fraction of relaxed clusters is $\sim44\%$ (considering all the 3240 galaxy clusters in $z\in[0,1.031]$). Note that our sample is not mass-complete at redshift $z>0$. Considering the offset parameters between BCGs and centroids, the relaxed cluster fractions for $MCC$ and $J$ thresholds are, respectively, $49\%$ and $47\%$ for $\Delta_{BCG-C_X}$ and $57\%$ and $47\%$ for $\Delta_{BCG-C_y}$. The difference between $MCC$ and $J$ for the $y$ offset is due to the different thresholds suggested by the two summary statistics. The peak of $MCC$ distribution is flatter compared to $J$, as shown in Fig.~\ref{fig:MCC_J}, suggesting a higher threshold and, consequently, a larger fraction of relaxed clusters. The different number of relaxed clusters from dynamical to morphological parameters comes from contamination of hybrid and disturbed clusters, as explained in Sec.~\ref{ssec:res_Xy}. The clusters that are identified as relaxed both with dynamical indicators and $M$ are just $31$-$33\%$ of the total, while the fraction of false detection (the non-relaxed clusters by dynamical indicators identified as relaxed by $M$ with these thresholds) is close to $13$-$16\%$. For completeness, the number of relaxed clusters from dynamical state indicators that are not recognised as such by $M$ is $\sim11$-$14\%$. The samples in Table~\ref{tab:relax_fraction} differ both in redshift and mass ranges, but our results are still in agreement with \citet{Rossetti2016} or \citet{Lopes2018} if we compare sub-samples with similar redshift range and medians of the previously cited work. Considering only {\sc The Three Hundred} clusters with $z\leq0.116$, the fraction of relaxed clusters from $M_{X}$ is $45$-$49\%$ ($48$-$49\%$ for $M_y$), while for $z\leq0.304$ it is $\sim49$-$50\%$ ($44\%$ for $MCC$ threshold in $M_X$). For clusters in $z\leq0.557$ or $z\in[0.116,0.663]$ we have a percentage of $49$-$50\%$ ($45\%$ again from $MCC$ threshold for $M_X$) of relaxed clusters.

Morphological parameters have also been studied intensively with numerical simulations. \citet{Pinkney1996} examined several statistical tests for substructure detection in optical maps, while \citet{Rasia2013} review the performances of X-ray based morphological parameters. On mock SZ maps, \citet{Cialone2018} study the cluster morphology using morphological parameters originally defined for X-ray, while \citet{Capalbo2020} apply a Zernike polynomial decomposition for the morphological analysis. Several works take advantage of simulation to study the correlation between the mass bias or, more in general, the mass estimates and the cluster morphology \citep{Piffaretti2008, Rasia2012, Green2019, Barnes2020} or with other cluster properties as the mass accretion rate \citep{Rasia2014, Chen2019}, the correlation between centre offsets and gas velocity dispersion \citep{Li2018}, the ICM thermodynamical profiles \citep{Ruppin2019} and turbulence \citep{Valdarnini2019}. In a recent work by \citet{Cao2021}, the consistency of relaxed cluster fractions and the thresholding problem in the relaxation definitions are studied with simulated galaxy clusters taken from IllustrisTNG \citep{Marinacci2018}, BAHAMAS \citep{McCarthy2016} and MACSIS \citep{Barnes2016} simulation suites. They found that the effectiveness of a single relaxation threshold and the consistency of relaxed sub-samples from different parameters are limited due to the intrinsic scatter of the morphological parameters, numerical resolution and subgrid physics, dependency by redshift and mass or the arbitrary nature of relaxation threshold values. 

The introduction of combined parameters \citep{Rasia2013, Meneghetti2014, Lovisari2017, Cialone2018, Bartalucci2019}, like $M$, could contribute to reducing the variation in the fraction of relaxed galaxy clusters. Representing a weighted average of different clusters dynamical feature, the single definition of each used parameter lose importance in the choice of a threshold on $M$, avoiding non-trivial cross consistency check among different indicators. Furthermore, it gives a unique and continuous estimation of galaxy clusters regularity or relaxation, even if the performances of combined parameters must first be investigated with the advantage of numerical simulations, as was done previously in \citet{Rasia2013} and \citetalias{Cialone2018}. Concerning these two works, in this paper we test the performance of combined parameters for a larger sample of galaxy clusters, both in redshift and size (in \citet{Rasia2013} they use 60 Chandra-like images of 20 simulated galaxy clusters, while in \citetalias{Cialone2018} there are 258 clusters, studied in 4 redshift snapshots) and studied with a multi-wavelength approach.

\section{Conclusions} \label{sec:conclusion}

In the literature, there is no consensus, both in simulations and observation \citep[e.g.][]{Cui2017, Cao2021}, in how to divide clusters according to their dynamical state with both dynamical or morphological indicators. In this paper, we study the performance of these kinds of indicators using galaxy clusters from {\sc The Three Hundred} Project {\sc gadget-x} simulation. From each of the ten redshift snapshots between $z\in[0;1.031]$ that we used, we have extracted the 324 most massive central galaxy clusters for a total of 3240 objects with masses: $M_{500}=(0.15$-$17.58)\times10^{14}h^{-1}$M$_{\sun}$. From this sample, we estimated and compared the dynamical state of clusters using three different relaxation classifications. For this purpose, we used in total five 3D indicators commonly used in simulations: the total sub-halo and the most massive substructure fractions in mass $f_s$ and $f_{s,mm}$, the viral ratio $\eta$, the centre of mass offset $\Delta_r$ and the relaxation parameter $\chi$. For the same clusters, synthetic multi-wavelength images have been produced to characterise the morphological state and test the performance of the morphological indicators. In particular, we used the same six indicators, plus a combination of them, already adopted in \citetalias{Cialone2018}, for our X-ray and SZ (described by the Comptonization parameter $y$) mock images. From the optical maps, the positions of the BCGs were determined to infer the dynamical state from indicators based on offsets between BCGs and X-ray or $y$ peaks or centroids positions. Our findings can be summarised as follows:

\begin{description}

\item Considering the relaxation criteria from \citet{300}, \citetalias{Cialone2018}, and a new one introduced in this work, the fraction of relaxed clusters strongly depend on which dynamical indicators are used, on the discrimination thresholds, and on the selected volume in which parameters are calculated. No remarkable difference between $R_{200}$ and $R_{500}$ is present for \citet{300} criterion due to the influence of $\eta$ dynamical indicator, while a slight increase is obtained with \citetalias{Cialone2018} one, which is due to the suppression of redshift evolution of the dynamical state induced by $f_{s,mm}$. Instead, the redshift evolution and the volume dependence are recovered considering the dynamical classification of this work. The dynamical state, however, is better described by continuous indicators rather than classes. The introduction of the $\chi$ indicator by \citet{Haggar2020}, as a combination of dynamical indicators, has the advantage to combine the different dynamical property of the other parameters giving a single continuous indicator for the dynamical state to compare with morphological indicators.

\item As for the dynamical state, the morphology of galaxy clusters in X-ray and $y$ maps is better described by the continuous combined parameter $M$. After the tuning procedure of the six parameters which constitute $M$, this parameter works efficiently and with comparable results on the two maps: for $M_y$-$M_X$ relation a Spearman correlation of $\rho=0.80$ is present. Moreover, the best fit of the data is in agreement with the identity line. Regarding the link between morphology and dynamical state, $M$ shows a relatively strong correlation with the dynamical state parameter $\chi$, for which $\rho\sim-0.66$. Considering the dynamical classification, the major source of contamination on relaxed sub-sample is composed of hybrid clusters, as highlighted from the analysis of ROC curves. Considering two dichotomous tests between relaxed and disturbed clusters and relaxed against non-relaxed (disturbed plus hybrid) sub-samples, the ROC curves underline a decrease of performances of $M$ discrimination ability when hybrid objects are included in the test. The area under the ROC curve decreases by $\sim11\%$ using the non-relaxed sample, from $AUC\sim0.9$ to $\sim0.8$. Consequently, the rates of contamination from disturbed and hybrid clusters on relaxed sub-sample are different. From the $P$ merit function, the contamination is close to $P\sim45\%$ for hybrid and $\sim20\%$ for disturbed at $M\sim0$, while disturbed $P$ decreases faster than hybrid one for negative values of $M$.

\item For the offset parameters, the position of BCGs and X-ray or $y$ peaks are good tracers of the peak density of clusters: no remarkable differences are present in their distributions depending on the dynamical state. Therefore, the offset parameters between BCGs and X-ray, $y$ peaks are not efficient dynamical state parameters, with ROC curves close to the performance of a random guess classifier. Considering instead the offsets between BCGs and the centroids of $y$ or X-ray maps, the efficiency of these offset parameters are comparable to $M$, with $AUC\sim0.8$ for the binary test of relaxed with non-relaxed sub-samples. The correlation between $M$ and these offsets are high ($\rho\sim0.80$), while the correlation with the dynamical state is different for $y$ ($\rho=-0.63$) and X-ray ($\rho=-0.69$) centroids, but both relatively strong. Similar results are obtained if peaks are used instead of BCGs positions.

\item Considering the lack of consensus in the literature about the actual fraction of relaxed galaxy clusters in observation, our relaxed sub-sample is comparable with \citet{Rossetti2017} and \citet{Bartalucci2019} results, and with the fraction of \citet{Lopes2018} obtained by the offset with X-ray centroids and BCGs. To be not biased by an arbitrary choice of the threshold with which segregate relaxed from non-relaxed, we use two summary statistic related to ROC curves, the Youden's $J$ statistic and the Matthews correlation coefficient $MCC$. In particular, we select as thresholds the values that maximise these two scores. We obtain a median relaxed cluster fraction of $\sim49\%$ from $M$ or the offset between BCGs and X-ray centroids. Instead, the two scores return two different fractions for the BCG-$y$ centroid offset parameter: $47\%$ for $J$ and $57\%$ for $MCC$. This discrepancy is due to the relation between the dynamical state and the offset parameters. For the $y$ centroid, the $MCC$ peak is flatter than in the J parameter, suggesting a slightly larger threshold ($0.06$-$0.07R_{500}$ instead of $0.05R_{500}$). However, this corresponds to a variation of $10\%$ in the fraction of relaxed clusters, underling how problematic the thresholding problem could be for relaxation definition.
\end{description}

\section*{Acknowledgements}
This work has been made possible by {\sc The Three Hundred} Collaboration. The simulations used in this paper have been performed in the MareNostrum Supercomputer at the Barcelona Supercomputing Center, thanks to CPU time granted by the Red Espa\~nola de Supercomputaci\'on. As part of {\sc The Three Hundred} Project, this work has received financial support from the European Union's Horizon 2020 Research and Innovation Programme under the Marie Sklodowska-Curie grant agreement number 734374, the LACEGAL project.


MDP and FDL acknowledge support from Sapienza Università di Roma thanks to Progetti di Ricerca Medi 2019, prot. RM11916B7540DD8D. WC acknowledges supports from the European Research Council under grant number 670193 (the COSFORM project) and from the China Manned Space Program through its Space Application System. 
GY and AK acknowledge financial support from \textit{Ministerio de Ciencia, Innovación y Universidades / Fondo Europeo de DEsarrollo Regional} (Spain), under research grant PGC2018-094975-C21. 
We thank Veronica Biffi for the help in generating the maps and Giammarco Cialone for useful comments. AK further thanks The Mabels for shifting sands.
This research made use of several {\sc python} packages: {\sc numpy} \citep{numpy}, {\sc scipy} \citep{SciPy}, {\sc matplotlib} \citep{matplot}, {\sc seaborn} \citep{seaborn}, {\sc pandas} \citep{pandas}, {\sc astropy} \citep{astropy,astropy2}, {\sc photutils} \citep{photutils}.

\section*{Data Availability}

The data used in this paper is part of {\sc The Three Hundred} Project and can be accessed following the guidelines of the collaboration that can be found on the main website\footnote{\url{https://the300-project.org}} of the collaboration. The data specifically shown in this paper will be shared upon request to the authors.

\bibliographystyle{mnras}
\bibliography{biblio}

\appendix

\section{Roc analysis on morphological parameters}
\label{sec:A_ROC}

\begin{figure*}
	\includegraphics[width=\columnwidth]{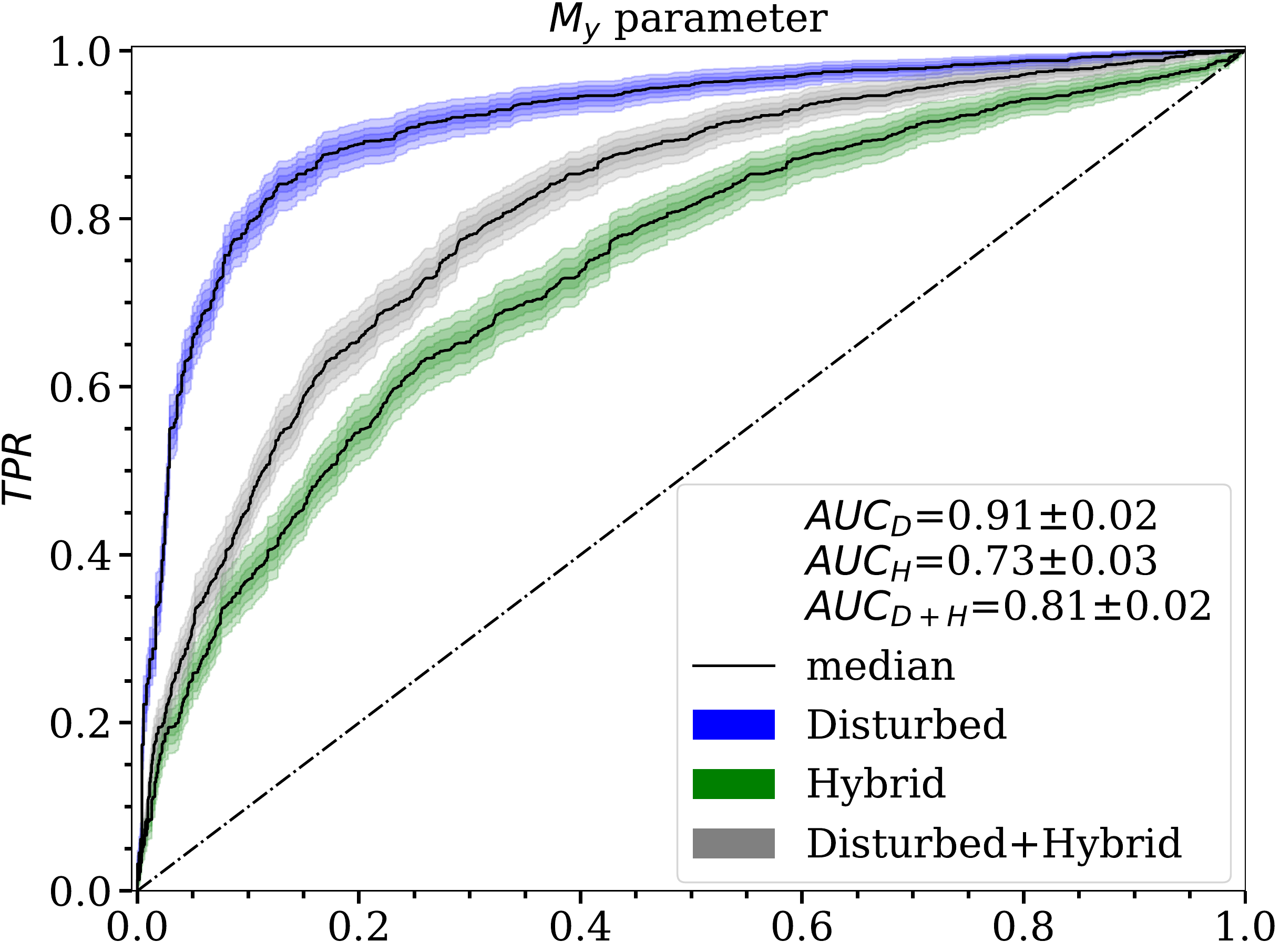}
	\includegraphics[width=\columnwidth]{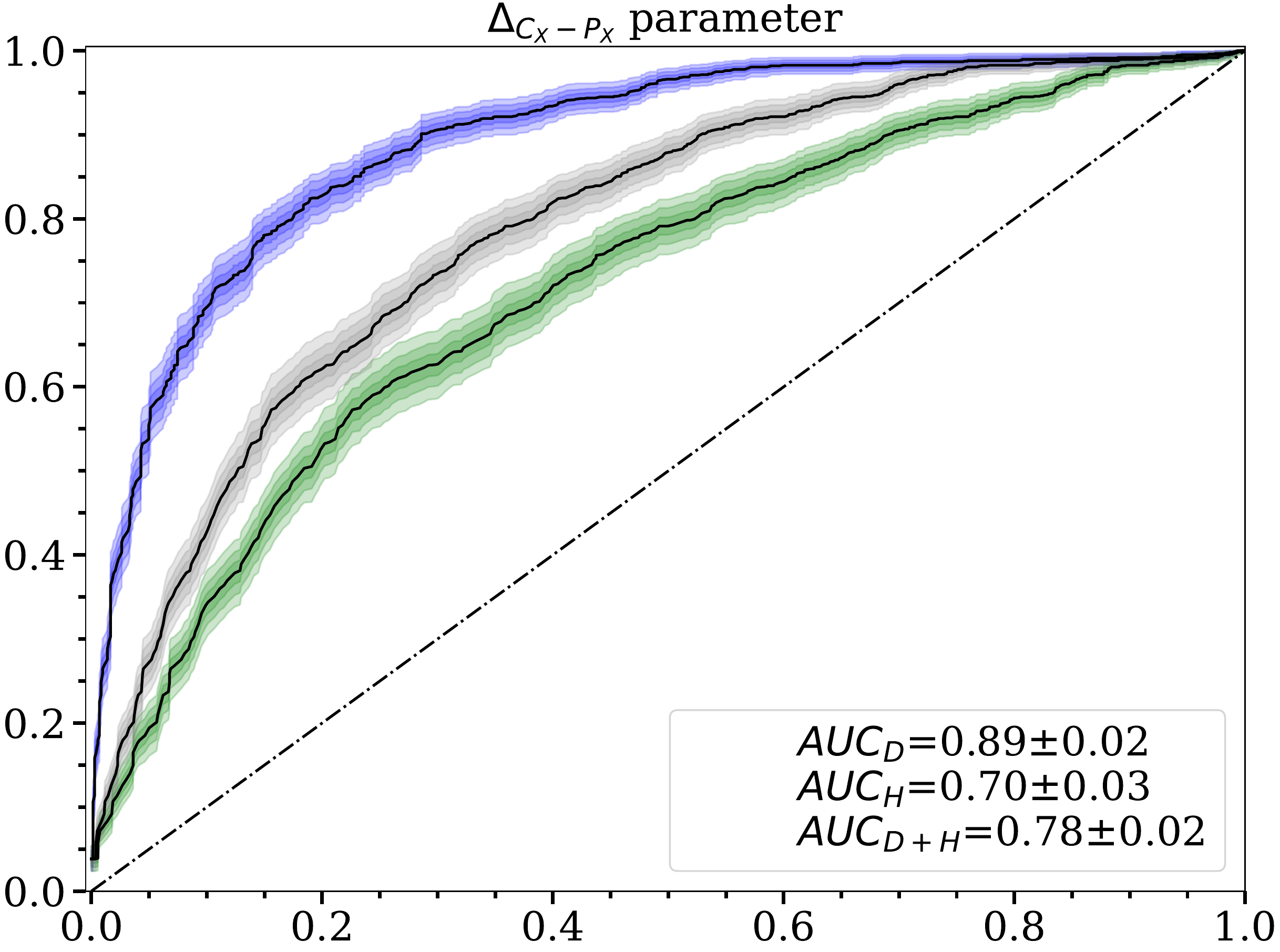}
	\includegraphics[width=\columnwidth]{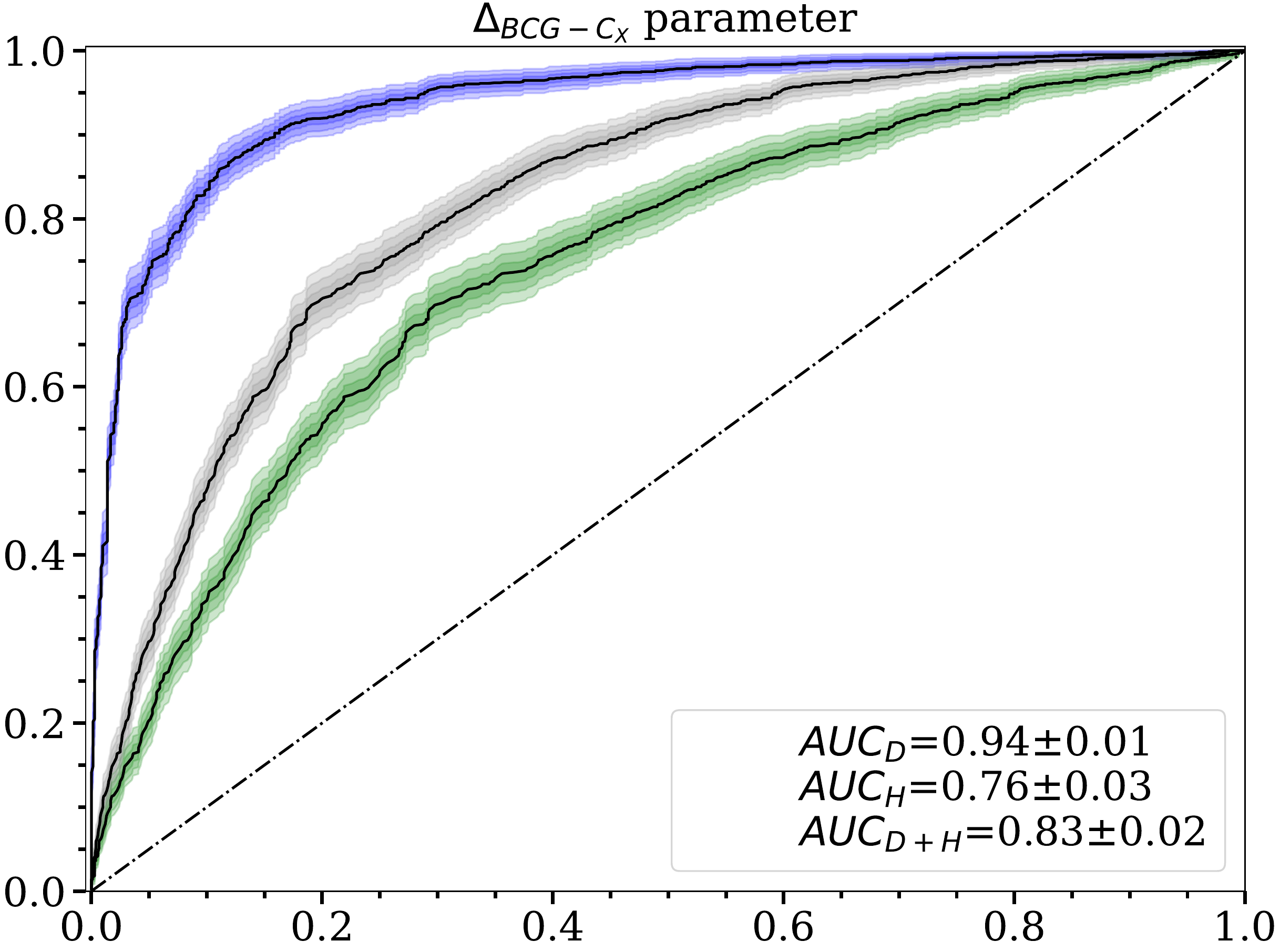}
	\includegraphics[width=\columnwidth]{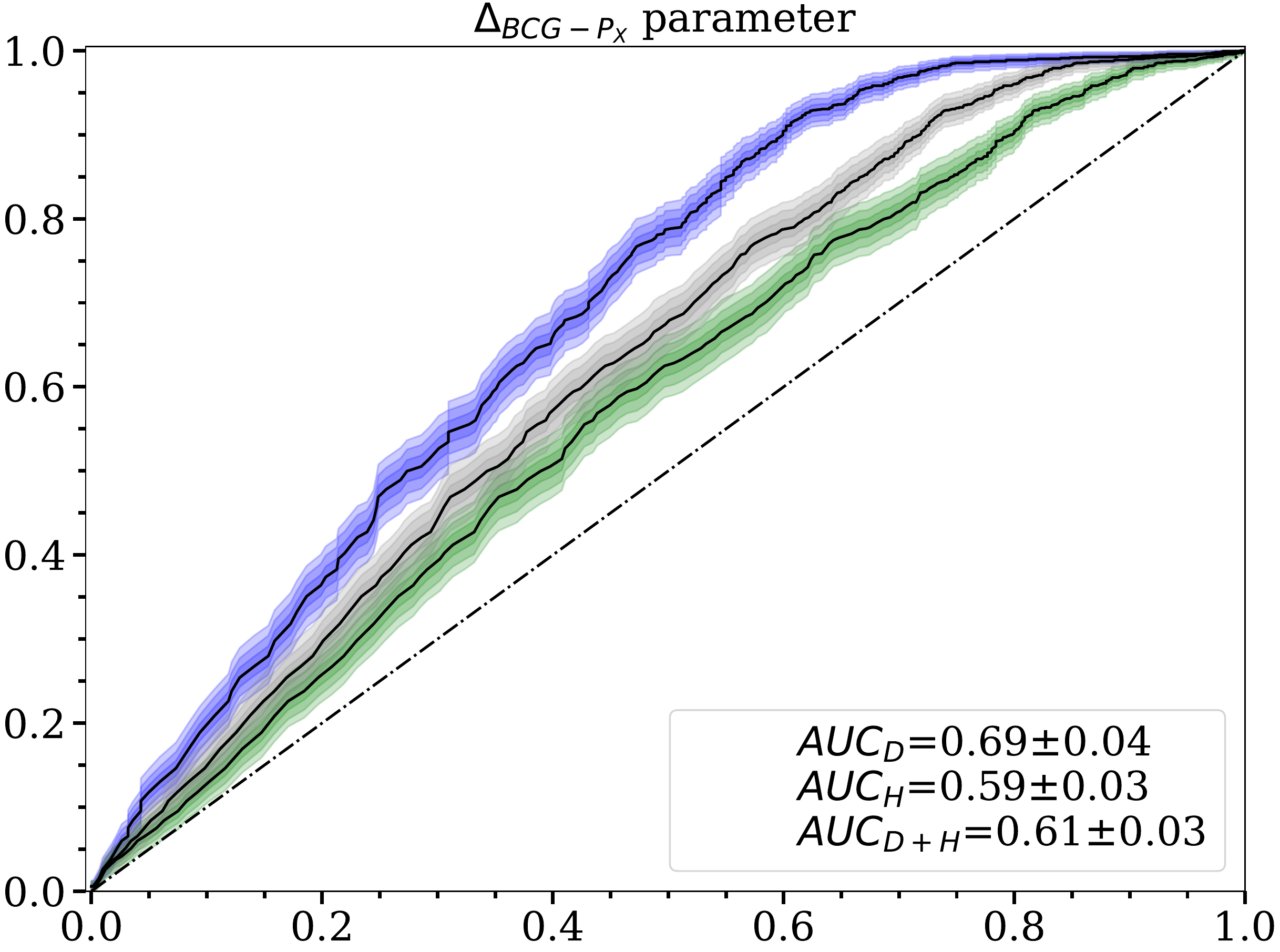}
    \caption{Median ROC curves considering all the 3240 galaxy clusters. Blue, green, and grey coloured areas correspond to the bootstrap $68$-$95$-$99\%$ confidence intervals for, respectively, the relaxed against only disturbed, hybrid, and disturbed plus hybrid ROC curves. The upper row shows the ROC curves for the $M_y$ parameter (left panel), and for the offset parameter between the X-ray peak and centroid (right). In the second row the difference of performance in the ROC curves, when the X-ray centroid (left panel) or peak (right) positions are considered for the offset from BCG as a morphological parameter. The median $AUC$ is shown in each legend of the panels for the three tests, with a $99\%$ confidence interval. The identity line in this plane corresponds to the performance of a random guess classifier.}
    \label{fig:ROC}
\end{figure*}

ROC curves, initially introduced as a method to characterise radar receivers, are now widely used in many scientific applications, as medicine \citep{Infantino2020} or machine learning techniques for dichotomous (or more) classifiers. The main advantage of ROC analysis is that it represents, graphically, the diagnostic ability of a test when an arbitrary threshold is varied. In our work, we decided to use ROC curves in order to illustrate the thresholding problem of the relaxation definition of galaxy clusters. As outlined in Sec.~\ref{ssec:2D_obs_comp}, or more in detail by \citet{Cao2021}, there is no consensus in the literature on which threshold to use for a given morphological parameter. This leads to different relaxation criteria that could be very restrictive or not. A ROC inspection of the diagnostic power could provide a criterion to select which are the best parameters to use in observation and give information about possible threshold on them, binding their definition to some summary statistic drawn from the curve, as has been done here with $AUC$, $MCC$, and $J$. 

The ROC curves for morphological parameters $M$ on $y$ maps and the offset between the BCG and the X-ray centroid ($\Delta_{BCG-C_{X}}$) are shown in the left column of Fig.~\ref{fig:ROC}. In all the panels, we consider in the tests all the 3240 clusters in the sample. The confidence intervals are computed by performing a bootstrap (with 4000 realisations) of the efficiency estimators. Similar results to the ones show in Fig.~\ref{fig:ROC} are also obtained for X-ray maps $M$ parameter and BCG-$y$ centroid offset parameter. The two morphological parameters are more efficient in separating the two extremes of the dynamical state. The disturbed ROC curves are always higher than hybrid ones, as the area under the curve: $AUC\gtrsim0.90$ for disturbed and $AUC\sim0.74$ when we consider only the hybrid in the test. As a result, if we are interested to extract a relaxed sub-sample, the possible contaminants consist mainly of hybrid clusters. Comparing the relaxed and non-relaxed objects, we have an intermediate performance with $AUC\sim0.82$. For the offset parameters, we show in the second row of Fig.~\ref{fig:ROC} the difference of performances when the peaks are used instead of centroids. Comparing the curves in the two panels, a sharp drop in the performance is evident when the peak is used: $AUC$ fells from $0.83$ (for $\Delta_{BCG-C_{X}}$, considering the relaxed versus the non-relaxed test) to $0.61$ ($\Delta_{BCG-P_{X}}$) and the ROC curves are closed to the identity line, that in this plane represents the performance of a random guess classifier. This lack of performances is related to X-ray or $y$ peaks positions, which are not good indicators of the dynamical state but are reliable tracers of the peak density of galaxy clusters. Therefore all the possible offsets between BCG, X-ray or $y$ peaks and peak density show no dependence from the dynamical state and have similar ROC curves to the one presents in Fig.~\ref{fig:ROC}. As a result, using the positions of the peaks or BCGs do not affect dramatically the results, as illustrated in the upper right panel of Fig.~\ref{fig:ROC}, where the offset $\Delta_{P_X-C_X}$ between the X-ray centroid and peak is shown. In this case, the $AUC$ is slightly lower than BCG ones: $AUC\sim0.89$ for disturbed and $AUC\sim0.78$ for non-relaxed classes.

\section{Convergence of efficiency indicators} \label{ssec:covergence}

\begin{figure}
	\includegraphics[width=\columnwidth]{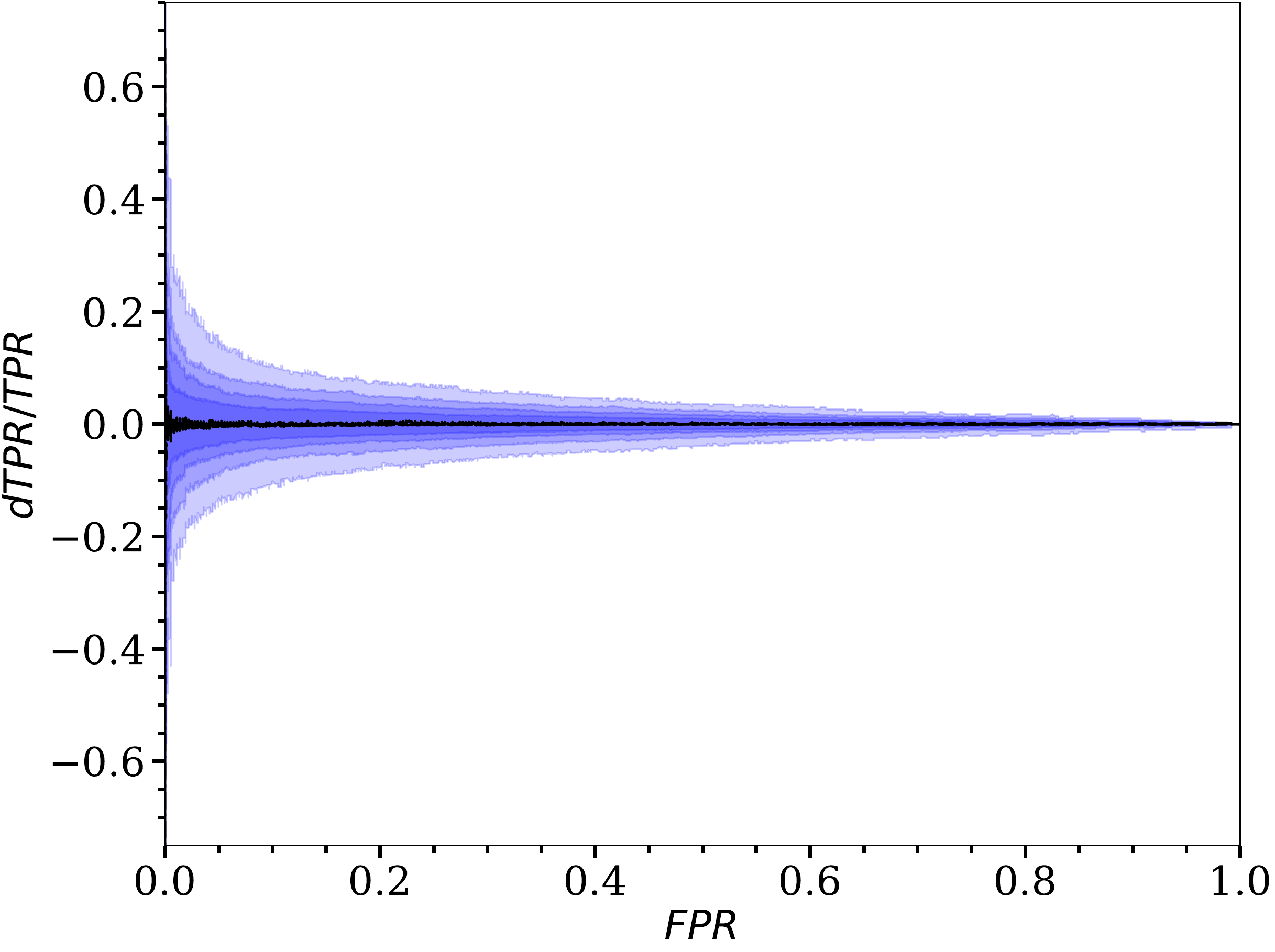}
    \caption{Relative deviation from the overall $TPR$ profile considering the four resamplings, for the $M_X$ morphological parameter. The coloured filled areas represent the $95\%$ dispersion of the data. The intensities of the colours are sorted in descending order concerning a reduction of 20, 40, 60, 80\% of the sample.}
    \label{fig:conv_MX}
\end{figure}

The sample analysed in this work consists of a large collection of simulated galaxy clusters, in a wide redshift range, compared to the typically observed sample, as shown by Table~\ref{tab:relax_fraction} in Sec.~\ref{ssec:2D_obs_comp}. To test if the results shown in this paper remain stable even for smaller samples, we estimate the dispersion of the efficiency parameters in cases where the sample is reduced by 20, 40, 60 and 80 per cent. For these resamplings, we realise 1000 realisations where the clusters are randomly selected, without replacement, but keeping the fraction of relaxed, hybrid and disturbed objects unchanged with respect to the overall sample. Considering the ROC curves for the $M_X$ morphological parameter, the deviations between the median of the realisations, relative to the overall profile of the $TPR$, are shown in Fig.~\ref{fig:conv_MX} as solid black lines. The dispersion of $TPR$ values is represented, in the same figure, with a blue filled area with upper and lower limits relative to the 2.3th and 97.7th percentiles of the data. The intensities of the colours in the figure are sorted in descending order, concerning the sample reduction. The darkest blue is associated with a reduction of 20\%, while the lighter with a reduction of 80\%. As expected, the dispersion becomes larger when the number of clusters reduces, but, statistically, the efficiency estimator converges to the overall profile. We obtain similar results for the other efficiency parameters as $FPR$, $MCC$, and $J$, and considering the other morphological parameters. For the efficiency parameters, the medians of the realisations converge to the overall profiles but with different dispersion depending on the parameter. The $FPR$ parameter has a dispersion generally larger than $TPR$, while $MCC$ and $J$ show similar dispersion between them. For the area under the ROC curve, the $AUC$ distributions have a median compatible with the overall value, with an increasing dispersion for smaller samples. As an example, for the $M_X$ parameter the median $AUC$ from all the resamplings converges to the overall value of $0.80$, but with an increasing dispersion from $\pm 0.02$ (reducing the sample of $20\%$) to $\pm 0.05$ ($80\%$). 


\bsp	
\label{lastpage}
\end{document}